# AutoGEEval++: A Multi-Level and Multi-Geospatial-Modality Automated Evaluation Framework for Large Language Models in Geospatial Code Generation on Google Earth Engine

Huayi Wu[a,b], Zhangxiao Shen[a], Shuyang Hou[a]*, Haoyue Jiao[c], Ziqi Liu[a], Lutong Xie[a], Chang Liu [a], Jianyuan Liang[a], Yaxian Qing[a], Xiaopu Zhang[a], Dehua Peng[d], Zhipeng Gui[d], Xuefeng Guan[a]

*a. State Key Laboratory of Information Engineering in Surveying, Mapping and Remote Sensing, Wuhan University, Wuhan, China

b. Collaborative Innovation Center of Geospatial Technology, Wuhan University, Wuhan, China

c. School of Resource and Environmental Sciences, Wuhan University, Wuhan, China

d. School of Remote Sensing and Information Engineering, Wuhan University, Wuhan, China

*Corresponding author: Shuyang Hou, email: whuhsy@whu.edu.cn

## Abstract

Geospatial code generation is emerging as a critical frontier in the integration of artificial intelligence and geo-scientific analysis. However, standardised automated evaluation tools for this task remain lacking. This study introduces AutoGEEval++, an enhanced evaluation framework built upon AutoGEEval. It is the first automated assessment framework tailored for large language models (LLMs) performing geospatial code generation on the Google Earth Engine (GEE) platform, featuring support for diverse geospatial data modalities and multiple levels of task complexity. Built upon the GEE Python API, the framework includes a benchmark dataset—AutoGEEval++-Bench—comprising 6,365 test cases across 26 geospatial data types and three task categories: unit test, combo test, and theme test. It integrates both a submission programme and a judge programme to enable a fully automated, end-to-end evaluation pipeline, from code generation to execution-based verification. The framework incorporates multi-dimensional metrics, including accuracy, resource consumption, run-time efficiency, and error types, balancing hallucination control with operational efficiency. It also supports boundary capability testing and error pattern analysis. Using this framework, we systematically evaluate 24 state-of-the-art LLMs as of June 2025, spanning general-purpose, reasoning-enhanced, code-centric, and geoscience-specific models. Experimental results reveal distinct performance, stability, and error characteristics across task types, model architectures, and deployment scenarios, validating the practical utility and scalability of AutoGEEval++ in the context of vertical-domain code generation. This study establishes the first standardized evaluation protocol and foundational resource suite for GEE-based LLM code generation, offering a unified benchmark for performance comparison and a methodological paradigm for systematic evaluation from natural language to domain-specific code—thus advancing the frontier of geospatial AI research.

**Keywords:** Geospatial Code Generation; Large Language Models; Google Earth Engine; Automated Evaluation

## 1. Introduction

General-purpose code refers to program instructions written in formal languages such as Python, C++, or Java, designed to control computer operations for specific tasks (Li et al., 2022; Popat and Starkey, 2019). The development of such code demands strict logical reasoning and substantial technical expertise (Jiang et al., 2024). In recent years, Transformer-based large language models (LLMs), including GPT-4o (Hurst et al., 2024), DeepSeek (Liu et al., 2024), Claude (Caruccio et al., 2024), LLaMA (Touvron et al., 2023), and Qwen2.5-Coder (Hui et al., 2024), have shown impressive capabilities in generating general-purpose code. These models enable users to create structured, executable code directly from natural language inputs, greatly reducing the technical barrier to programming (Zheng et al., 2023). At the same time, benchmark systems such as HumanEval (Li and Murr, 2024), MBPP (Yu et al., 2024), and LiveCodeBench (Jain et al., 2024)have facilitated the quantitative evaluation of LLMs across dimensions such as executability, accuracy, and efficiency. With the rise of intelligent applications, increasingly specialized data processing needs have given rise to domain-specific code (Sun et al., 2024). Platforms such as Bioconductor (for bioinformatics) (Gentleman et al., 2004)and QuantLib (for quantitative finance) (Varma and Virmani, 2016)encapsulate domain-specific logic and data structures within general-purpose languages, forming highly specialised programming frameworks that reflect a re-syntactisation and functional repurposing of the underlying language.

**1. Function Invocation Error**

**User Requirements to LLM**

I want to add two MODIS NDVI images and get the projection information of the resulting image using the Earth Engine Python API.

**Example of Erroneous Output by LLM**

```
import ee
ee.Initialize()
ndvi1 = ee.Image('MODIS/006/MOD13A2/20210101').select('NDVI')
ndvi2 = ee.Image('MODIS/006/MOD13A2/20210117').select('NDVI')
ndvi_sum = ndvi1 + ndvi2
ndvi_sum = ndvi1.add(ndvi2)
print(ndvi_sum.projection().getInfo())
```

**2. Object Type Confusion**

**User Requirements to LLM**

Create a three-band Sentinel-2 image, convert it to an array, and extract the first band from the array image using the Earth Engine Python API.

**Example of Erroneous Output by LLM**

```
import ee
ee.Initialize()
image = ee.Image([0.1, 0.2, 0.3]).rename(['B4', 'B3', 'B2'])
array_image = image.toArray()
first_band = array_image.select('B4')
first_band = array_image.arrayGet([0])
print(first_band.getInfo())
```

**3. Missing Filter Conditions**

**User Requirements to LLM**

Retrieve the first image of Landsat 8 Surface Reflectance data over Paris in June 2022 and clip it to the city area using the Earth Engine Python API.

**Example of Erroneous Output by LLM**

```
import ee
ee.Initialize()
roi = ee.Geometry.Point([2.3522, 48.8566])  # Paris
collection = ee.ImageCollection('LANDSAT/LC08/C02/T1_L2') \
    .filterBounds(roi)
collection = ee.ImageCollection('LANDSAT/LC08/C02/T1_L2') \
    .filterBounds(roi) \
    .filterDate('2022-06-01', '2022-06-30')
image = collection.first().clip(roi)
print(image.getInfo())
```

**4. Loop Structure Error**

**User Requirements to LLM**

Iterate through the first five images in a Landsat 9 collection over Tokyo and print their IDs using the Earth Engine Python API.

**Example of Erroneous Output by LLM**

```
import ee
ee.Initialize()
collection = ee.ImageCollection('LANDSAT/LC09/C02/T1_L2') \
    .filterBounds(ee.Geometry.Point([139.6917, 35.6895])) \
    .limit(5)
for i in range(5):
    image = collection.get(i)
    print(image.id())
def get_image_id(image):
    return ee.Image(image).id()
image_ids = collection.map(get_image_id)
print(image_ids.getInfo())
```

**5. Band Mapping Error**

**User Requirements to LLM**

Select the shortwave infrared band from a Landsat 8 Surface Reflectance image using the Earth Engine Python API.

**Example of Erroneous Output by LLM**

```
import ee
ee.Initialize()
image = ee.Image('LANDSAT/LC08/C02/T1_L2/LC08_044034_20210610')
swir_band = image.select('B5')
swir_band = image.select('B6')
```

**6. Type Mismatch Error**

**User Requirements to LLM**

Apply a constant temperature value to a defined region in Brazil using the Earth Engine Python API.

**Example of Erroneous Output by LLM**

```
import ee
ee.Initialize()
temperature = ee.Image.constant(30)
region = ee.Geometry.Rectangle([-54, -15, -53, -14])
result = temperature.add(region)
result = temperature.clip(region)
print(result.getInfo())
```

**7. Invalid Type Conversion**

**User Requirements to LLM**

Convert a filtered Landsat 8 image collection from July 2021 into a single image for display using the Earth Engine Python API.

**Example of Erroneous Output by LLM**

```
import ee
ee.Initialize()
collection = ee.ImageCollection('LANDSAT/LC08/C02/T1_TOA') \
    .filterDate('2021-07-01', '2021-07-31')
image = ee.Image(collection)
image = collection.mosaic()
print(image.getInfo())
```

**8. Missing Parameter**

**User Requirements to LLM**

Use the Earth Engine Python API to calculate the average reflectance value from a Landsat 9 image using 'reduceRegion'.

**Example of Erroneous Output by LLM**

```
import ee
ee.Initialize()
image = ee.Image('LANDSAT/LC09/C02/T1_TOA/LC09_030034_20220701')
mean = image.reduceRegion(reducer=ee.Reducer.mean())
mean = image.reduceRegion(
    reducer=ee.Reducer.mean(),
    geometry=image.geometry(),
    scale=30
)
print(mean.getInfo())
```

**Figure 1. Representative Error Types in LLM-Generated GEE Code.** Each example includes the error type, the user's geospatial prompt, and the resulting erroneous code. Errors are highlighted in red; corrected versions are shown in green.

In the geosciences, the widespread deployment of high-resolution remote sensing imagery and crowd-sourced sensors has resulted in geospatial data that are large in volume, highly diverse, and strongly correlated in space and time (Huang et al., 2024). Traditional general-purpose code often encounters bottlenecks when processing complex data types such as hyperspectral imagery, GeoJSON, digital elevation models (DEM), multi-temporal remote sensing data, synthetic aperture radar (SAR), vector, and raster data—especially during data ingestion, parallel computation, and indexing (Tamiminia et al., 2020). To address these challenges, cloud-based geospatial computing platforms—most notably Google Earth Engine (GEE)—have emerged as dominant tools for remote sensing and spatial analysis (Velastegui-Montoya et al., 2023). GEE provides both a JavaScript web interface and a Python API, offering extensive built-in functions (e.g. ‘ee.*’, ‘Export.*’) that enable users to programmatically perform cloud removal, radiometric correction, index calculation, and land change detection. This supports a code-centric paradigm for geospatial analysis (Yang et al., 2022). Compared with GUI-driven GIS software such as ArcGIS and QGIS, GEE offers greater automation and reusability, promoting the scalable dissemination and reuse of analytical workflows across users, regions, and disciplines (Amani et al., 2020; Zhao et al., 2021).

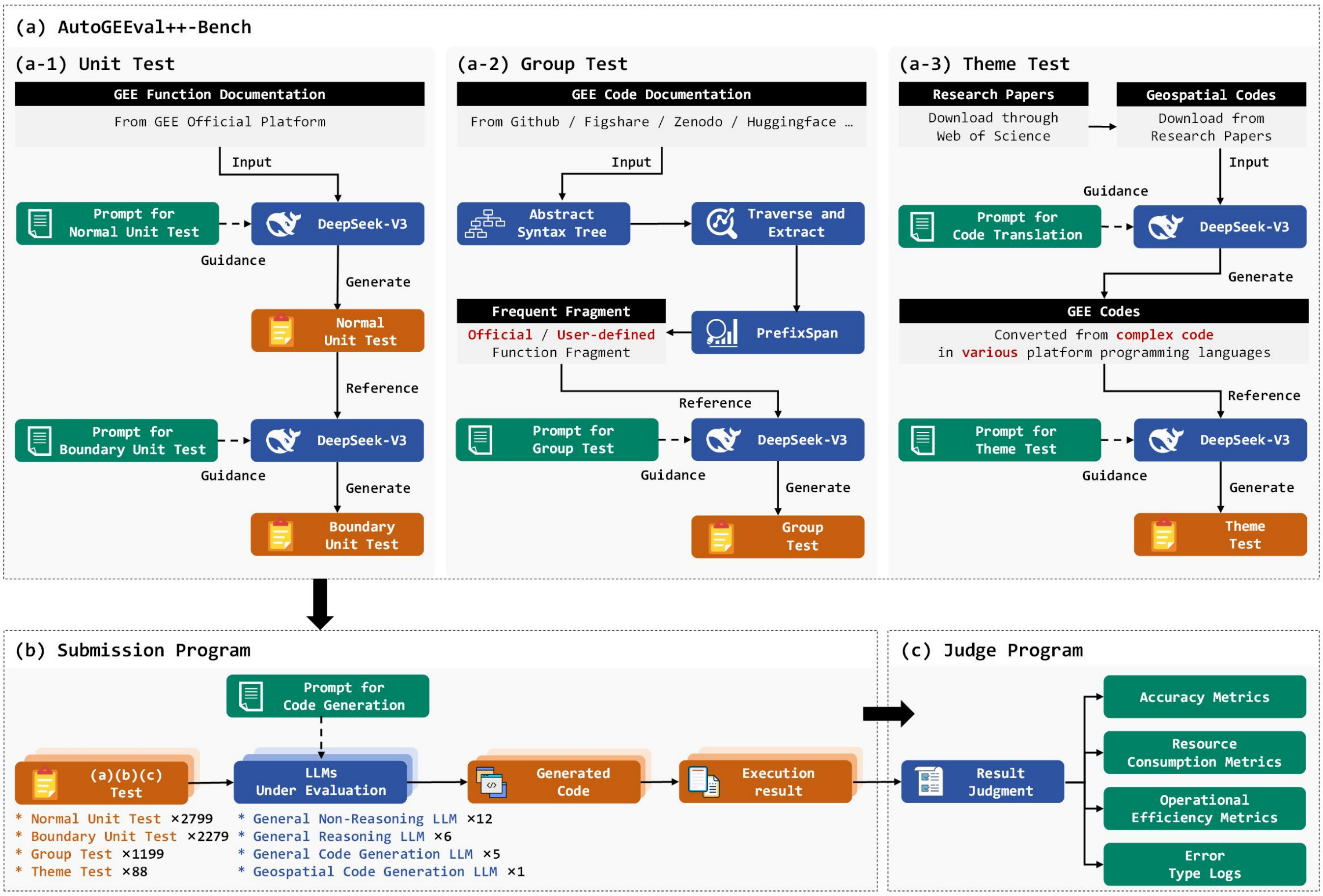

**Figure 2. AutoGEEval++ Framework Overview.** Black indicates source and intermediate documents; green denotes prompts or metric types; blue represents LLMs and intermediate processes; orange marks test queries and outputs. Solid arrows indicate the main workflow; dashed arrows show prompt flows guiding LLM execution.

Despite its flexibility, programming in GEE requires more than basic coding skills. It demands expertise in geospatial analysis, comprehension of core objects such as ‘ee.Image’ and ‘ee.FeatureCollection’, familiarity with remote sensing datasets like Landsat and MODIS, and knowledge of spatial concepts including coordinate systems and map projections (Hou et al., 2025a). Moreover, users must possess the ability to work with multimodal geospatial data, rendering the skill threshold significantly higher than that for general-purpose programming (Hou et al., 2025c). Although LLMs can be adapted for GEE code generation, the relatively low proportion of geoscientific content in their training data often leads to code hallucinations—outputs that appear plausible in meaning but contain syntactic errors or flawed logic, resulting in execution failures or distorted results (Gramacki et al., 2024). Figure 1 presents typical examples of such errors, including faulty API calls, object confusion, missing conditionals, and generation of empty map layers, all of which impair both executability and output reliability. This underscores the need for a systematic evaluation framework to define the operational boundaries of LLMs in geospatial domains and to guide future improvements in domain-adapted model performance.

Several preliminary efforts have explored evaluation mechanisms for geospatial code generation, including GeoCode-Bench (Hou et al., 2024b), GeoCode-Eval (Hou et al., 2025c), the GeoSpatial-Code-LLMs Dataset (Gramacki et al., 2024), and AutoGEEval (Hou et al., 2025b). GeoCode-Bench and GeoCode-Eval rely on manual evaluation, which incurs high cost and suffers from low reproducibility. The GeoSpatial-Code-LLMs Dataset introduces automation but remains limited in scale (40 samples) and lacks sufficient data type coverage. Although AutoGEEval extends to over 1,000 test cases, it focuses solely on single-function calls, without addressing boundary conditions or complex task scenarios. Furthermore, most of these studies are based on LLMs released prior to April 2025, rendering their findings outdated for assessing current mainstream model performance.

To this end, we propose AutoGEEval++, an enhanced evaluation framework for LLMs generating geospatial code on the GEE platform. Building upon AutoGEEval, the framework introduces support for multiple geospatial data modalities and hierarchical task complexities. Implemented via the GEE Python API, AutoGEEval++ is designed for local execution environments, enabling precise exception capture and automated feedback—capabilities that are difficult to realize via GEE's JavaScript interface due to browser-based constraints. As illustrated in Figure 2, the framework comprises three components: the AutoGEEval++-Bench test suite (Fig. 2a), the Submission program (Fig. 2b), and the JUDGE program (Fig. 2c). The benchmark includes 5,078 unit tests (2,799 general and 2,279 boundary), 1,199 combo tests, and 88 theme tests, covering 26 core GEE data types. General tests are derived from official documentation, boundary tests simulate edge-case parameter configurations, and combo tests are synthesized from over 290,000 real-world scripts using AST parsing, PrefixSpan mining, and LLM-based reconstruction. Theme tests are curated from 946 code examples extracted from 93 WoS-indexed publications, refined through human validation and model translation. Each test case specifies six elements: function declaration, reference implementation, parameter list, expected output type, output path, and ground truth. The Submission program generates and executes code based on task prompts, while the JUDGE program evaluates output accuracy and logs resource consumption, execution time, and error categories. Leveraging this framework, we assess 24 mainstream LLMs released before June 2025, including 12 general-purpose, 6 reasoning-augmented, 5 code-specialized, and 1 geoscience-specific model. The evaluation provides a comprehensive analysis of model performance and optimization potential in geospatial code generation scenarios.

The main contributions of this study are summarized as follows:

- Proposing AutoGEEval++, a multi-modal, multi-level automated evaluation framework for unit, combo, and theme tasks in geospatial code generation;
- Open-sourcing the AutoGEEval++-Bench benchmark, which covers 26 geospatial data types and includes 6,365 test cases;
- Systematically evaluating 24 LLMs (as of June 2025) on GEE-based geospatial code generation, with metrics covering stability, accuracy, resource consumption, runtime efficiency, boundary test pass rate, and error type distribution—thereby providing a comprehensive landscape of model capabilities and optimization needs.

The remainder of this paper is organized as follows. Section 2 reviews geospatial code, code generation tasks, and LLM-based evaluation research. Section 3 introduces the construction of the AutoGEEval++-Bench benchmark. Section 4 describes the evaluation pipeline and the design of the submission and grading modules. Section 5 presents and discusses the experimental results. Section 6 concludes the study with a summary of findings, a discussion of limitations, and an outlook on future research directions.

## 2. Related work

### 2.1. Geospatial code

"Geospatial code" is a domain-specific extension of general-purpose programming, tailored for the processing, analysis, and visualization of geospatial data in geoscientific contexts (Hou et al., 2024a). It is distinct from terms such as "geocoding (Zandbergen, 2008)" or "geospatial encoding (Mai et al., 2022)," which refer to data identifier transformation rather than analytical computation. The concept dates back to the 1963 Canadian Geographic Information System (CGIS) (Gierman and MacDonald, 1982) and was further advanced in the 1980s by the adoption of modular toolchains and AML scripting in GRASS GIS (Neteler et al., 2012) and ArcInfo (Morehouse, 1992). Since the 1990s, the convergence between general-purpose languages and GIS tools has accelerated. GRASS began supporting Perl and Python (Zambelli et al., 2013), and spatial libraries such as GDAL (Qin et al., 2014), GeoPandas (Bediroglu, 2025), Turf.js (Netek et al., 2023), Mapping Toolbox (Parsons et al., 2013), and sf emerged across platforms like Python, JavaScript, MATLAB, and R—contributing to the development of a local, code-driven geospatial analysis ecosystem. In the 2010s, APIs like ArcPy (Toms and O'Beirne, 2017)and PyQGIS (Isaya Ndossi and Avdan, 2016) promoted the shift from GUI-based desktop GIS to programmatic control. The emergence of GEE further catalyzed the transition toward cloud-based geospatial coding. With support for JavaScript and Python interfaces, GEE enables remote access and parallel processing of remote sensing data, driving the adoption of the "geospatial code as analysis" paradigm (Hou et al., 2024b).

### 2.2. Code Generation

Early code generation relied on heuristic rules and template-based systems (Srivastava et al., 2013), such as pattern matching in GCC and Lex/Yacc (Hou et al., 2025b). While these methods were computationally efficient, they l (Li et al., 2022) exhibited near-human-level algorithm-solving performance. Models such as DeepSeek-Coder (Guo et al., 2024) and Qwen2.5-Coder (Hui et al., 2024) have since refined and extended the technical landscape.

Geospatial code generation emerged more recently and initially relied on template-based systems, such as GEE's parameterized script generator and GPT-3-based code completion within ArcGIS Pro, both of which were limited to their respective platforms (Nowak et al., 2020). In October 2024, two systematic evaluation studies formally defined geospatial code generation as a standalone research task and extended the NL2Code paradigm to "NL2GeospatialCode," thereby establishing its theoretical foundations (Gramacki et al., 2024; Hou et al., 2024b). Follow-up research has introduced prompt chaining via chain-of-prompt (CoP) methods (Hou et al., 2024a), knowledge base construction with Geo-FuB (Hou et al., 2024c) and GEE-Ops (Hou et al., 2025a), fine-tuned models like GeoCode-GPT (Hou et al., 2025c), and intelligent systems such as GeoAgent (Yu and Peuquet, 2009), ShapefileGPT (Lin et al., 2024), and GIS Copilot (Akinboyewa et al., 2025)—collectively advancing the field from automation toward intelligent geospatial code generation and contributing to the formation of a dedicated research ecosystem.

### 2.3. Code Generation Evaluation

The evaluation of general-purpose code has shifted from syntax-based rules and static analysis toward metrics emphasizing execution correctness and semantic understanding. Benchmarks such as HumanEval (Li and Murr, 2024) and MBPP (Yu et al., 2024) assess generated code based on natural language descriptions and corresponding test cases, and are widely used for standard function-level tasks. However, these benchmarks fall short in addressing the multimodal inputs, task heterogeneity, and platform dependencies inherent in geospatial code generation.

Geospatial code evaluation has only recently begun receiving systematic attention. Frameworks such as GeoCode-

Bench (Hou et al., 2024b)and GeoCode-Eval (Gramacki et al., 2024) rely heavily on expert manual annotation, which introduces subjectivity, limits reproducibility, and lacks structured function-level testing. The GeoSpatial-Code-LLMs Dataset (Gramacki et al., 2024) proposes an automated evaluation approach, but its limited sample size and omission of key data types—such as remote sensing imagery—restrict its utility. Furthermore, it does not adequately model platform diversity or task complexity. The AutoGEEval (Hou et al., 2025b) framework fills this gap by introducing the first automated evaluation system for geospatial code targeting the GEE. It enables end-to-end unit-level testing with multimodal inputs and execution-based verification, using a standardized benchmark of 1,325 test cases. This forms a preliminary unified evaluation protocol for the field. However, AutoGEEval remains constrained to basic function calls, lacks support for function composition, real-world scenarios, and boundary conditions, and only evaluates models released prior to April 2025. To address these shortcomings, we propose AutoGEEval++, which extends the original framework by incorporating increased task complexity, broader data diversity, and updated model coverage. AutoGEEval++ represents the first multimodal, multi-tier, end-to-end automated evaluation framework for geospatial code generation, establishing a foundation for standardized benchmarking and future model advancement.

## 3. AutoGEEval++-Bench

This chapter presents the three types of test tasks in AutoGEEval++-Bench—unit tests, combo tests, and theme tests—covering their GEE data type coverage, task structure, definition, construction methodology, and final outcomes.

### 3.1. GEE Data Types

The output types $T_i$ in AutoGEEval++-Bench cover all 26 core data types of the GEE platform (see Table 1), encompassing a wide range of structures such as imagery, vector, raster, temporal, classifiers, kernels, and projection metadata.

**Table 1. Classification of Output Data Types in AutoGEEval++-Bench**

| Id | Output_type | Description |
|---|---|---|
| 1 | ee.Array | Multi-dimensional array for numbers and pixels |
| 2 | ee.ArrayImage | Image constructed from multidimensional arrays |
| 3 | ee.Blob | Binary large object storage (e.g., files/models) |
| 4 | ee.Classifier | Machine learning classifier object |
| 5 | ee.Clusterer | Clustering algorithm processor |
| 6 | ee.ConfusionMatrix | Confusion matrix of classification results |
| 7 | ee.Date | Date and time format data |
| 8 | ee.DateRange | Object representing a range of dates |
| 9 | ee.Dictionary | Key-value data structure |
| 10 | ee.Element | Fundamental unit of a geographic feature |
| 11 | ee.ErrorMargin | Statistical object for error margins |
| 12 | ee.Feature | Single feature with properties and shape |
| 13 | ee.FeatureCollection | Collection of geographic features |
| 14 | ee.Filter | Object representing data filtering conditions |
| 15 | ee.Geometry | Geometric shapes (point, line, polygon, etc.) |
| 16 | ee.Image | Single raster image data |
| 17 | ee.ImageCollection | Collection of image data objects |
| 18 | ee.Join | Method for joining datasets |
| 19 | ee.Kernel | Convolution kernel for spatial analysis |
| 20 | ee.List | Ordered list data structure |
| 21 | ee.Number | Numeric data |

| Id | Output_type | Description |
|---|---|---|
| 22 | ee.PixelType | Pixel type definition |
| 23 | ee.Projection | Coordinate system projection information |
| 24 | ee.Reducer | Aggregation and reduction functions |
| 25 | ee.String | String-type data |
| 26 | BOOL | Boolean logic value (True/False) |

### 3.2. Structural Design

In AutoGEEval++-Bench, each test case $q_i \in Q = \{q_1, q_2, ..., q_n\}$ is defined as a six-tuple:

$$q_i = (H_i, R_i, P_i, T_i, O_i, A_i) \quad (1)$$

The components are defined as follows:

- $H_i$: Function declaration, including the function name, parameter signature, and description, used to construct the prompt.
- $R_i$: Reference code generated by DeepSeek-V3 and manually validated; hidden during testing and used to generate the ground-truth output.
- $P_i$: Parameter list serving as input to the function.
- $T_i$: Output type (e.g., number, boolean, dictionary, image layer), used to call the corresponding evaluation module in the JUDGE program.
- $O_i$: Output path where the generated code stores its execution result, from which the JUDGE program reads for verification.
- $A_i$: Ground-truth answer produced by executing the reference code with given parameters, serving as the benchmark for automated evaluation.

AutoGEEval++-Bench uses YAML (YAML Ain't Markup Language) to define test cases, offering high human readability and machine interpretability (Měchura, 2023). Compared to JSON and XML, YAML supports comments and features a concise syntax, making it suitable for expressing nested structures like function definitions, parameter settings, and expected outputs. Its strong compatibility with Python allows direct conversion to dictionary objects, facilitating integration into function calls, prompt generation, and result evaluation workflows.

### 3.3. Task Definition

AutoGEEval++-Bench categorizes test tasks into three levels based on capability: Unit tests focus on single-function evaluation to assess the model's comprehension and invocation of the GEE API. These tests primarily use normal inputs, with some inclusion of abnormal values to evaluate boundary robustness, serving as the foundation for basic capability assessment. Combo tests examine the model's ability to compose multiple functions, constructed from real-world user-defined function combinations. Common structures are identified via pattern mining, with the goal of evaluating data flow construction and parameter propagation. Theme tests target typical application scenarios such as image classification, vector clipping, and temporal change detection. These require the model to interpret natural language task descriptions, invoke multiple functions, handle multimodal inputs, and generate executable analytical pipelines—representing the highest complexity and most realistic evaluation of a model's comprehensive performance.

The evaluation pipeline of AutoGEEval++ is jointly executed by the Submission program and Judge program. The Submission program extracts the function declaration $H_i$and constructs a prompt to guide model $M$ in generating the corresponding function body code:

$$c_i = M(Prompt(H_i)) \tag{2}$$

Then, the system injects the parameter list $P_i$ into the generated code and executes it locally, saving the result to path $O_i$:

$$\hat{y}_i = Run(c_i, P_i) \rightarrow O_i \tag{3}$$

Finally, the Judge program reads the model output $\hat{y}_i$ from $O_i$, nvokes the appropriate evaluation module based on the output type $T_i$ , and compares it against the reference answer $A_i$ to produce the final evaluation result:

$$Eval(q_i) = \begin{cases} 1, & if\ Judge(\hat{y}_i, A_i, T_i) = True \\ 0, & otherwise \end{cases} \tag{4}$$

### 3.4. Unit Test Construction

Unit tests are divided into general unit tests and boundary unit tests, as shown in Figure 3.

#### 3.4.1. General Unit Tests

General unit tests were constructed based on the Client Libraries section of the official GEE Reference documentation, covering 1,374 function entries. Each function page provides the function name, description, parameter and return type specifications, and example code (Figure 4). Prior to test generation, the research team manually validated the usability of each function, excluding 43 deprecated entries due to version changes. This resulted in 1,325 functions serving as the foundation for testing. The information on each function page was structured into JSON format. Using a predefined prompt template (Figure 5), DeepSeek-V3 was employed to automatically generate test queries. The selection of DeepSeek-V3 was motivated by its leading accuracy in the AutoGEEval benchmark. As the generated code was manually reviewed and only used to streamline test construction—and given that the final evaluation is based on execution success rather than code quality—this model choice does not introduce bias.

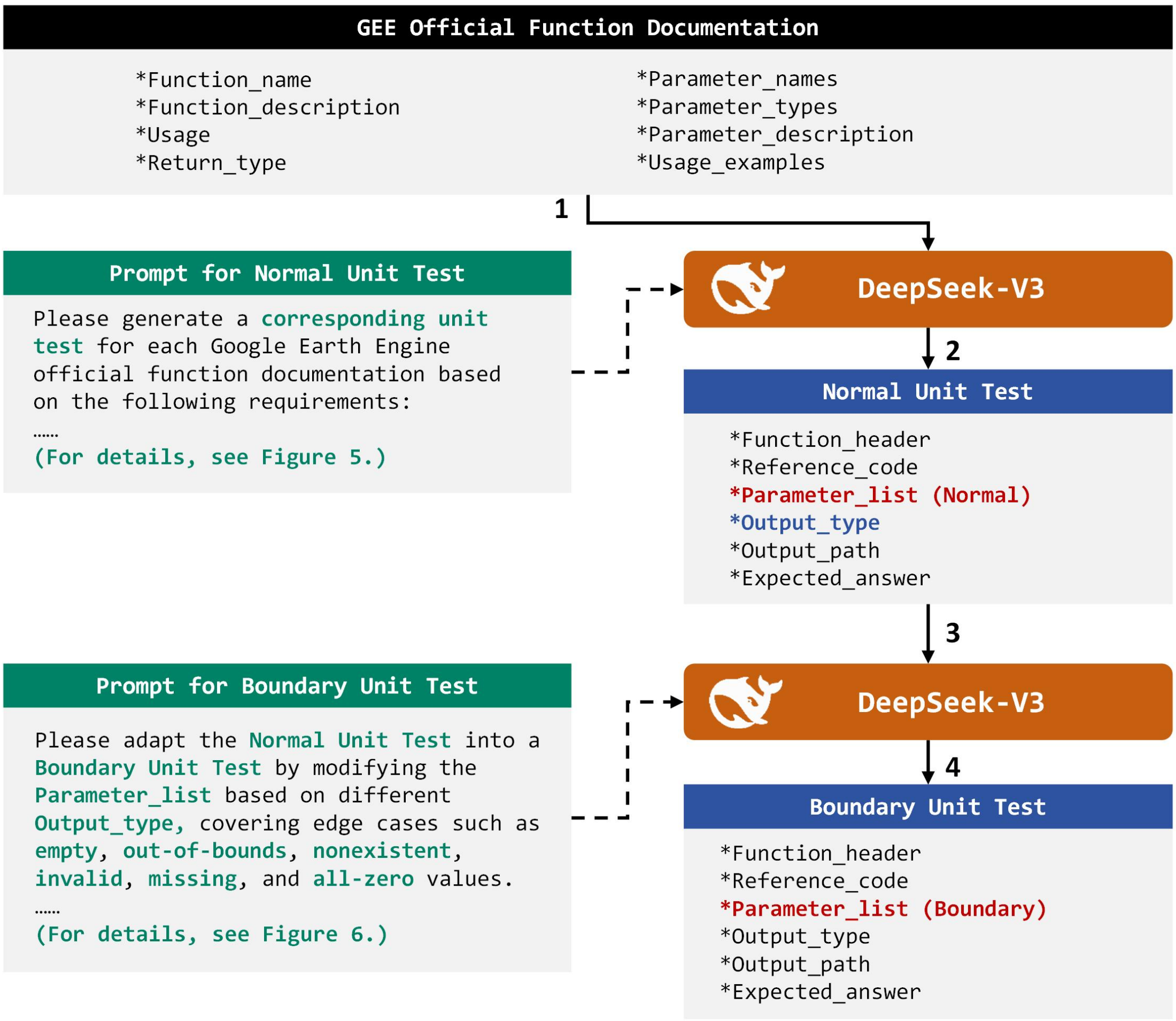

**Figure 3. Unit Test Construction Method.** Red highlights key differences between the two unit test types; blue indicates criteria for boundary test construction; green marks keywords in the prompts. The prompts shown are refined versions; full versions are provided in Figures 5 and 6.

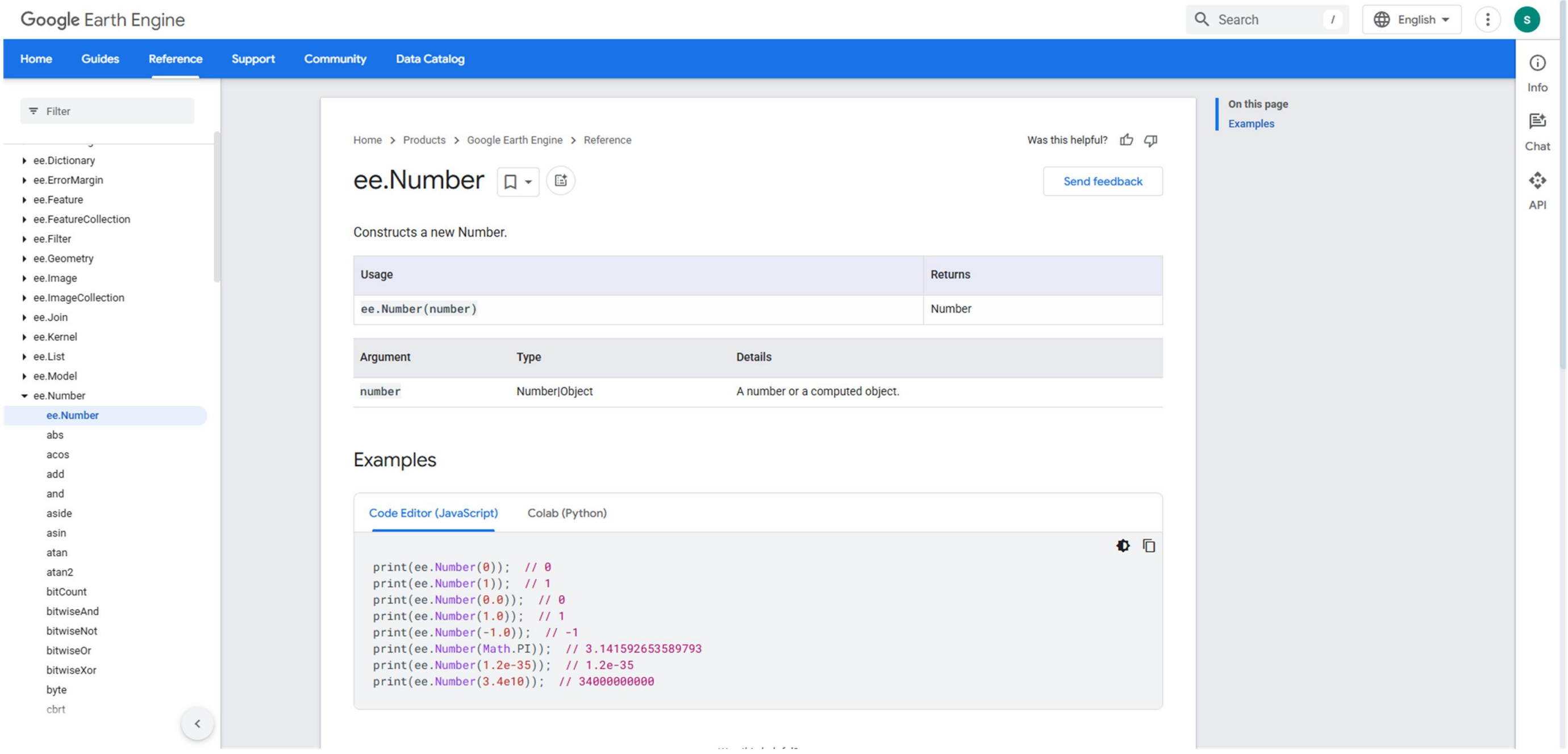

**Figure 4. Example Page from GEE API Reference.** The left sidebar allows for selecting different functions, while the main interface displays the relevant information for the selected function.

**Promt for Normal Unit Test**

```
## Task Description
You need to generate standard test code and configuration file entries for a given Google Earth Engine (GEE) Python API operator.
Each operator will have two parts: the standard code and the test cases in the configuration file.

### Input
1. **Operator Name**: Name of the operator
2. **Explanation**: The explanation of the operator about what it does
3. **Parameter List**: List of parameters with their types and descriptions. For example, `image` (ee.Image): The input image
4. **Return Type**: The return type of the operator

### Output
1. **Standard Code**: Define a function that uses the given operator and returns the result.
The function name should be (Data Type+ operator name + Task). For example, `ee.Image.NormalizedDifference`->`imageNormalizedDifferenceTask`.
2. **Test Cases in Configuration File**: Include multiple test cases, each with parameters, expected answer path, and output type.

### GEE objects in params
1.If the parameter is an GEE object(e.g. ee.Image, ee.Number, etc), use the following format in the configuration file to return the object with python:
param_name: !python |
    def get_ee_object():
        import ee
        ee.Initialize()
        # then get and return the wanted object
2.Notice that some operators may require specific GEE objects as input. e.g. 'ee.Array.CholoskyDecomposition' requires a positive definite ee.Array matrix.

### Output Type
1. The output type can be one of the following:
GEE objects:
"ee.Image", "ee.FeatureCollection", "ee.Number", "ee.List", "ee.Dictionary", "ee.Geometry", "ee.Array", "ee.ImageArray"
Python objects:
"str", "int", "float", "bool", "list", "dict", "NoneType"
2. You can use other types if needed.

### Expected answer
1. The value of the "expected_answer" field in the configuration file MUST be the path to the file containing the expected output.
2. The file name should be (function name + "_testcase" + testcase_number), file type should be .npy for images and arrays,
 .geojson for geometry or feature objects, .txt for other types.

### Note
1. The function should just include ONE operator and return the result. They are used for automatic testing.
2. If the output is a GEE object, do NOT perform getInfo() function. Just return the object.
3. Use the given operator for your answer, do NOT use other methods or operators to solve the task.
4. Any import statements, initialization statements or example usages are NOT needed.
5. Do NOT add any explanation.

### Operator Information
Here is the operator information:
```

**Figure 5. Prompt for normal unit test construction.**

### 3.4.2. Boundary Unit Tests

Boundary unit tests extend general unit tests by incorporating abnormal parameter configurations to assess model fault tolerance. These tests preserve the function declaration, reference code, and output type while generating abnormal parameter lists tailored to the semantic features of the output type $T_i$. The goal is to evaluate the model's logical consistency and execution stability under atypical inputs.

AutoGEEval++-Bench defines five anomaly categories: (1) null values (e.g., None or empty objects), (2) overflow (extreme numerical values), (3) missing inputs (omitted required parameters), (4) invalid inputs (type or structural mismatches such as incorrect geometries or malformed strings), and (5) all-zero data (e.g., images where all pixels equal zero). Table 2 summarizes the anomaly types and their applicable data categories, while Figure 6 provides a representative prompt used in boundary test generation.

**Promt for Boundary Unit Test**

```
## Task Description
You are tasked with creating YAML-formatted test cases for a Python function that wraps a Google Earth Engine (GEE) algorithm.
You are required to add **1-2** boundary test cases (with boundary_test: true) to the existing set of test cases, without altering their structure or
indentation.

1. You are given the function signature, docstring, and executable code.
2. You are also given the existing test cases in YAML format.
Each test case includes parameters (`params`), where GEE special objects (e.g., `ee.Image`) are defined using a `!python` tag that returns a GEE object.
Scalar parameters like `sigma` and `threshold` are also included.
Each test case also contains:
- `expected_answer`: A filename (.npy for arrays/images, .geojson for geometry and features, and .txt for other types).
- `out_type`: Expected output type (e.g., `ee.Image`, `bool`).
- `boundary_test`: Boolean indicating whether this is a boundary test case (`true` or `false`).

### Examples of boundary test cases include, but are not limited to:
1. Invalid geometry (e.g., empty geometry, coordinates out of bounds)
2. Empty images (e.g., from empty collections or masked-out regions)
3. Non-existent assets (e.g., invalid image IDs)
4. Invalid dates (e.g., malformed date strings)
5. Array/list index out of bounds or empty lists
6. Invalid parameter types or values
7. Missing required arguments
8. Numerical overflow or division by zero
9. Invalid projections or coordinate reference systems (CRS)

### Requirements
1. Add 1-2 new boundary test cases below the existing ones. If 2 boundary test cases already exist, do NOT add more.
2. Do NOT modify the structure or indentation of the existing test cases. Do NOT change or output the function code.

The function code:
    {code_content}
The existing test cases:
    {config_content}

Please provide the new test cases in YAML format, ensuring that the indentation and structure are consistent with the existing ones.
Return ONLY the YAML content (including the original test cases), with NO additional text or explanation.
"""
```

**Figure 6. Prompt for boundary unit test construction.**

**Table 2. Abnormal Input Design in Boundary Unit Tests**

| Exception | GEE Data Type | Example Abnormal Input or Condition |
|---|---|---|
| Null | ee.String | "" (empty string) |
| | ee.List | [] (empty list) |
| | ee.Array | ee.Array([]) (empty array) |
| | ee.Geometry | ee.Geometry([]) (empty geometry) |
| | ee.Image | Empty image constructed via ee.Image([]) |
| | ee.Number | None or uninitialized variable |
| | ee.ImageCollection | ee.ImageCollection([]) |
| | ee.FeatureCollection | ee.FeatureCollection([]) |
| Overflow | ee.Number | ee.Number(1e308) or ee.Number(1000).int8() → value truncated |
| | ee.Array | Array with values beyond allowed index or range |

| Exception | GEE Data Type | Example Abnormal Input or Condition |
|---|---|---|
| Missing | General Function Parameters | Calling function without required args, e.g. ee.Date() with no input |
| Invalid | ee.Geometry | Malformed GeoJSON or invalid coordinates |
| | ee.String | Non-string input or special characters (e.g. object instead of string) |
| | ee.Projection | Unrecognized CRS string, e.g. "EPSG:INVALID" |
| | ee.Dictionary | Dictionary with wrong key types or nested structure issues |
| | ee.Date | Illegal date format, e.g. '2025-05-31' |
| All-zero | ee.Image | Image filled with all-zero pixels |
| | ee.Array | All elements are zero, e.g. ee.Array([0, 0, 0]) |
| | ee.List | List like [0, 0, 0] with no meaningful content |
| | ee.Number | 0 used in sensitive context, e.g. as divisor |

### 3.5. Combination Tests Construction

Combination tests are constructed from a corpus of 295,943 user-authored GEE scripts sourced from GitHub, Zenodo, and HuggingFace, spanning the period from September 2015 to September 2024, with file sizes ranging from 1 to 533 KB. These scripts contain two main structural elements: standard API functions defined in the official GEE Reference (termed "official functions") and user-defined functions composed of multiple official calls, which exhibit high reusability and modularity.

The objective is to systematically mine high-frequency function compositions from this corpus and construct representative procedural patterns with sequential semantics. These patterns serve as the foundation for assessing LLMs' ability to generate coherent multi-function workflows. The construction process comprises three key stages: (1) building and traversing abstract syntax trees (ASTs), (2) mining function compositions, and (3) generating test items, as illustrated in Figure 7.

#### 3.5.1. AST Construction and Traversal

Due to the variability in user coding practices, rule-based methods are insufficient for accurately distinguishing between user-defined and official functions. To address this, all collected scripts are parsed into ASTs and serialized into JSON format, enabling structured analysis and systematic function extraction. In the AST representation, syntactic elements are encoded as specific fields, while nesting and function invocation relationships are explicitly captured in the hierarchical structure. Non-semantic content such as comments is automatically excluded, and scripts with syntax errors are discarded. A depth-first traversal algorithm is designed to extract function call paths and classify their types. The algorithm maintains four data structures: "processed_nodes" to avoid redundant visits, "function_stack" to track scope, "processed_functions" to record defined functions, and "operators" to store call sequences.

During traversal of the Python AST, it identifies FunctionDef nodes (user-defined functions), inserts [FUNCTION_START] and [FUNCTION_END] markers, and recursively parses their bodies. All function calls appear as "Call" nodes. For each, the func field is recursively expanded to reconstruct the full path by unraveling nested "Attribute" and Name nodes (e.g., "ee.Image", "image.select", "geometry.centroid"). Calls within a function body are prefixed with [function], while those in the main program body are not. Notably, invocations of user-defined functions themselves are not extracted as independent paths; only their internal logic is included.

The final output is an ordered sequence of function call paths, where [FUNCTION_START] and [FUNCTION_END]

enclose user-defined compositions and the rest represent official function compositions. This process yields 322,710 user-defined function fragments and 425,819 official ones, serving as input for downstream combination pattern mining. The full pipeline is illustrated in Figure 8.

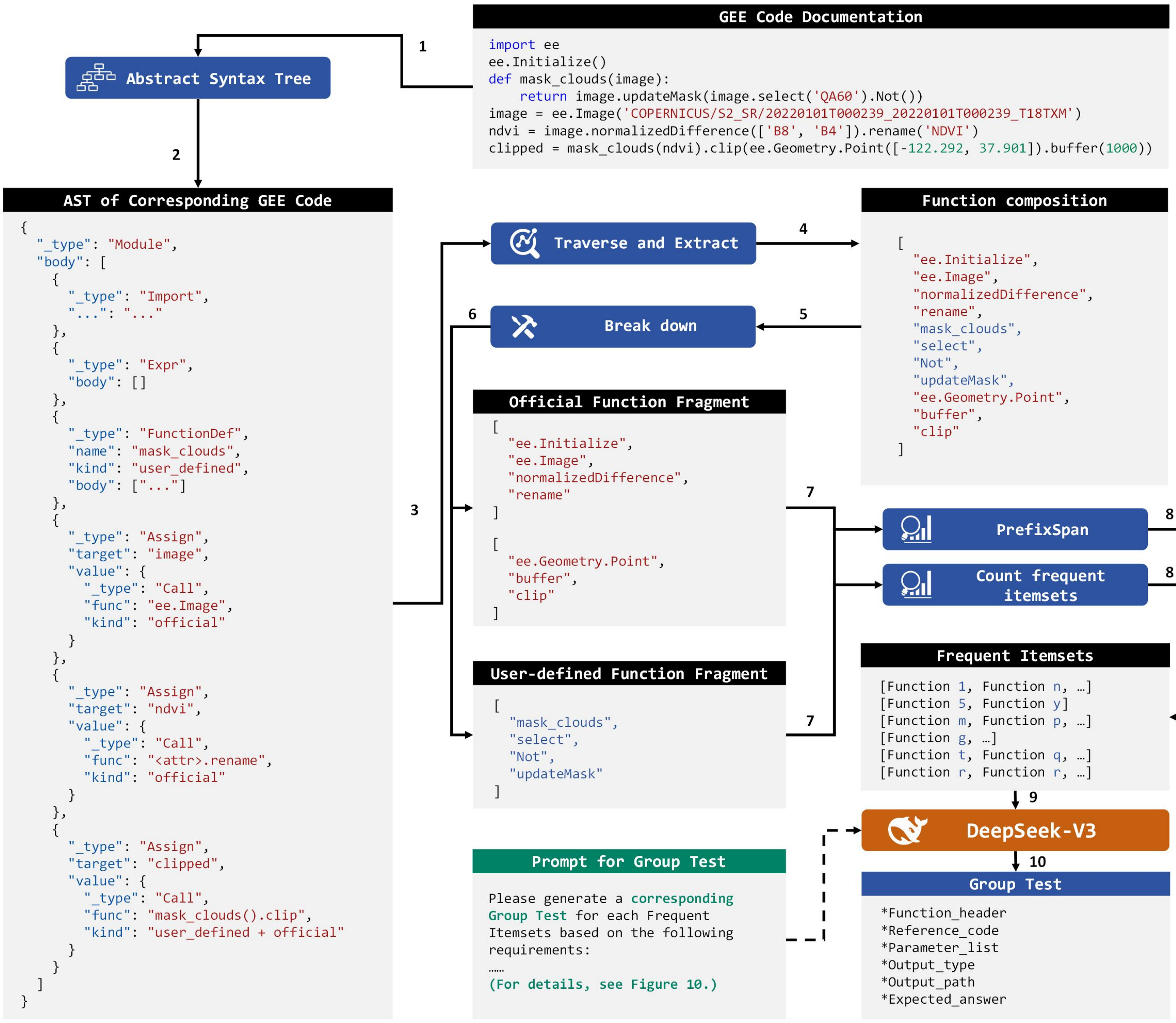

**Figure 7. Group Test Construction Method.**

### 3.5.2. Frequent Pattern Mining

For the extracted official function combinations and user-defined function combinations, this study adopts differentiated strategies to perform frequent pattern mining. For all sequences of official function combinations, this study applies the PrefixSpan algorithm to mine frequent subsequences. PrefixSpan is a sequential pattern mining method that considers the order of elements. Its core idea is to recursively divide the original sequence database into multiple projected subsets and mine frequent sequences starting with a specific prefix in each subset. In this study, the database of combined function sequences is defined as:

$$D = \{S_1, S_2, \dots, S_n\},\ S_i = \langle o_1, o_2, \dots, o_m \rangle \tag{5}$$

Where $o_j$ represents the $j-$th geospatial function. The goal of the PrefixSpan algorithm is to extract all frequent sequence pattern sets that meet a minimum support threshold $\theta$:

$$F = \{P | P \sqsubseteq S_i, \forall S_i \in D, and\ sup(P) \geq \theta\} \tag{6}$$

Where $P \sqsubseteq S_i$ indicates that sequence $P$ is a subsequence of $S_i$, and $sup(P)$ represents the support of $P$ in the database.

Specifically, the algorithm first identifies all frequent items of length 1 as prefixes, then constructs projected databases based on these prefixes, and further recursively extracts frequent patterns from their suffix sequences. Compared to traditional sequential mining algorithms (e.g., Apriori or FP-Growth), PrefixSpan effectively preserves the order information of operations in the sequence and avoids generating a large number of redundant candidates, offering higher efficiency and scalability.

For user-defined function combination sequences, in addition to using PrefixSpan to extract structured patterns, this study also directly counts the occurrence frequencies of all user-defined function fragments. Since user-defined function combinations are operation patterns actively constructed by users in practical scenarios, their frequency distribution has direct representativeness and evaluation value. Finally, the three types of results—mining results of official function combinations, mining results of user-defined function combinations, and frequency statistics of user-defined function combinations—are merged and deduplicated to form a unified data basis for constructing the combination test set.

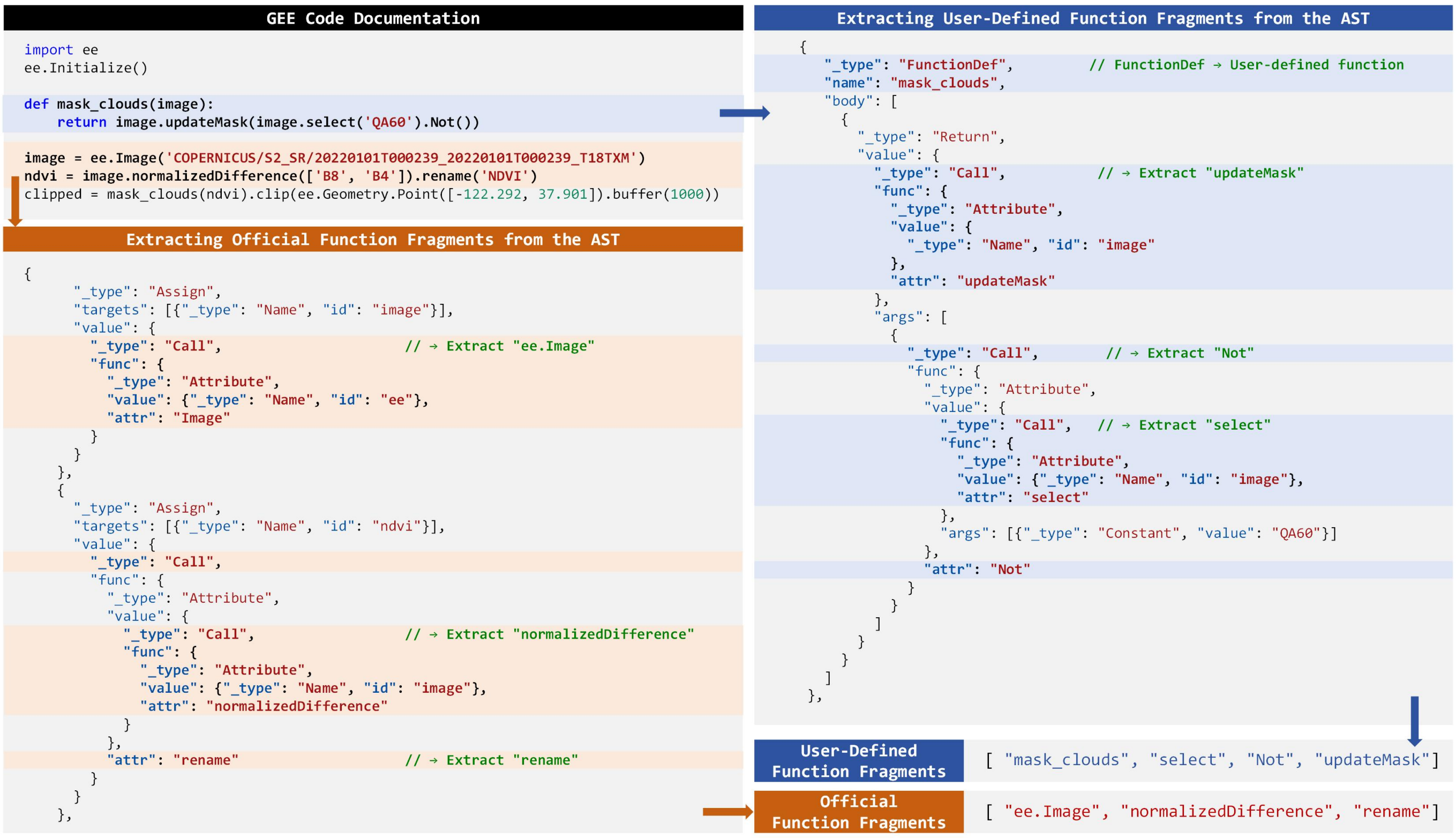

**Figure 8. Extracting Function Compositions from GEE Code AST.** Only selected JSON-formatted AST fragments are shown to illustrate the traversal-based extraction process. Orange highlights indicate official compositions; blue highlights denote user-defined ones. Key extraction segments are emphasized.

To optimize the construction of combination tests, this study conducted comparative experiments on the minimum support parameter θ for the PrefixSpan mining results of both official and user-defined function combinations (see Table 3), and also counted the frequency distribution of user-defined combinations (see Figure 9). Taking into account the number of sequences, average length, and the cost of test construction, the support threshold for official combinations was finally set to 0.10, resulting in 531 frequent combinations; for user-defined combinations, the threshold was set to 0.04, extracting 484 frequent combinations. In the frequency statistics, the inflection point of the distribution curve (150 occurrences) was used as the threshold, yielding 236 high-frequency combinations. After merging and deduplication of the three types of combinations, a total of 1,199 unique combination patterns were constructed as the basis for the combination test question bank.

**Table 3. Comparison of Minimum Support Thresholds in PrefixSpan-Based Pattern Mining.** Support indicates the minimum support threshold; Sequences refers to the number of mined operator composition patterns under each threshold; Average Length denotes the average number of operators per pattern. ✓ marks the threshold selected for final pattern extraction.

| Category | Support | Sequences | Average Length | Select? |
|---|---|---|---|---|
| User-Defined Function | 0.05 | 9101 | 4.75 | |
| | 0.06 | 4192 | 4.23 | |
| | 0.07 | 2166 | 3.78 | |
| | 0.08 | 1284 | 3.5 | |
| | 0.09 | 789 | 3.29 | |
| | **0.1** | **531** | **3.14** | √ |
| | 0.11 | 370 | 2.99 | |
| | 0.12 | 254 | 2.87 | |
| | 0.13 | 194 | 2.77 | |
| | 0.14 | 140 | 2.68 | |
| | 0.15 | 107 | 2.59 | |
| | 0.16 | 84 | 2.58 | |
| | 0.17 | 71 | 2.46 | |
| | 0.18 | 59 | 2.42 | |
| | 0.19 | 48 | 2.38 | |
| | 0.2 | 35 | 2.37 | |
| Official Function | 0.01 | 23223 | 6.04 | |
| | 0.02 | 2119 | 4.34 | |
| | 0.03 | 932 | 4.31 | |
| | **0.04** | **484** | **3.92** | √ |
| | 0.05 | 402 | 4.06 | |
| | 0.06 | 255 | 3.98 | |
| | 0.07 | 238 | 4.08 | |
| | 0.08 | 131 | 3.59 | |
| | 0.09 | 116 | 3.66 | |
| | 0.1 | 111 | 3.73 | |
| | 0.11 | 108 | 3.78 | |
| | 0.12 | 106 | 3.8 | |
| | 0.13 | 106 | 3.8 | |
| | 0.14 | 106 | 3.8 | |
| | 0.15 | 43 | 2.93 | |

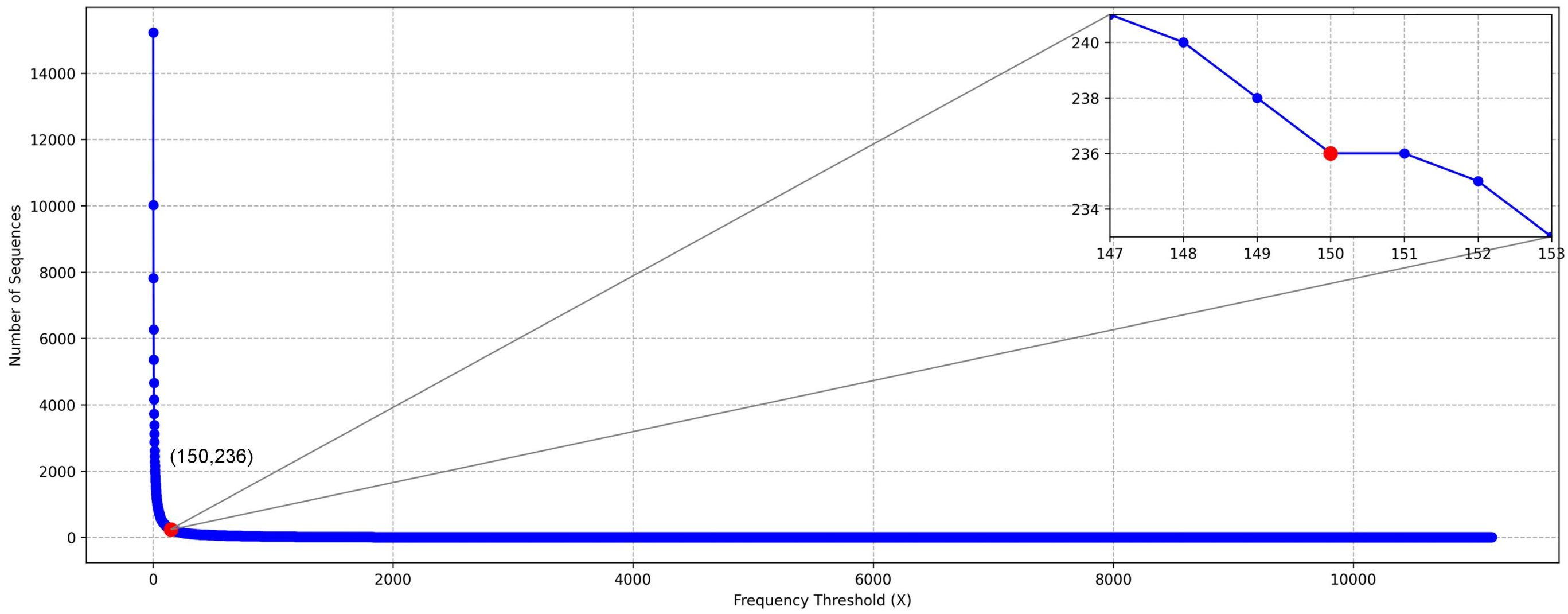

**Figure 9. Frequency Distribution of User-Defined Function Combinations.** The x-axis shows combination frequency; the y-axis indicates the number of combinations with frequency $\geq$x. The red point marks the elbow, identified using the maximum distance method, where the decline in frequency notably slows—highlighting the rarity and significance of high-frequency patterns.

### 3.5.3. Test Question Construction

For each group of frequent function combinations, the framework sequentially retrieves the definitions of each function from the GEE official API as described in Section 3.4.1, and concatenates them according to their original call order. A structured prompt is then constructed based on this order to guide DeepSeek-V3 to automatically read the function documentation and generate the corresponding combination test questions. A schematic of the prompt template is shown in Figure 10.

**Promt for Group Test**

```
## Task Description
You need to generate test code and configuration file entries for given Google Earth Engine (GEE) Python API operators.
Each operator will have two parts: the standard code and the test cases in the configuration file.

### Input
**Operators**: Name of the operator list with order, indicating a regular usage in GEE for spatiotemporal analysis.

### Output
1. **Standard Code**: Define a function that uses the given operators and returns the result.
The function name should be clear. It should have a docstring that explains the function and its parameters.
2. **Test Cases in Configuration File**: Include multiple test cases, each with parameters, expected answer path, and output type.

### Output Type
1. The output type can be one of the following:
GEE objects:
"ee.Image", "ee.FeatureCollection", "ee.Number", "ee.List", "ee.Dictionary", "ee.Geometry", "ee.Array", "ee.ImageArray"
Python objects:
"str", "int", "float", "bool", "list", "dict"
2. You can use other types if needed, but there MUST be ONE output.

### Expected answer
1. The value of the "expected_answer" field in the configuration file MUST be the path to the file containing the expected output.
2. The file name should be (function name + "_testcase" + testcase_number), file type should be .npy for images and
arrays, .geojson for geometry or feature objects, .txt for other types.

### Note
1. The tasks are used for automated testing, so they should be clear and concise.
2. Use the given operators and order for the task, but you can add additional operators or delete some of them if needed.
3. Do NOT use a whole Image to reduce computation, clip it into a smaller region first.
4. If the output is a GEE object, do NOT perform getInfo() function. Just return the object.
5. Any import statements or initialization statements are NOT needed.
6. Do NOT add any explanation or example usages.

### Operators
Here are the operators:
#### Example Input Operators
- **Operators**: ['addBands', 'select', 'map']
```

**Figure 10. Prompt for group test construction.**

### 3.6. Theme Test Construction

Theme test tasks are constructed based on peer-reviewed literature published since 2021 in the fields of geoscience and artificial intelligence, as shown in Figure 11. This study retrieved 93 relevant papers from the Web of Science using the query TS=("Large Language Model" OR LLM) AND TS=("GIS" OR "Geospatial Analysis" OR "Geospatial Code Generation" OR GEO). To construct representative evaluation tasks, the source code attached in the papers was collected from platforms such as GitHub, Figshare, HuggingFace, and PapersWithCode, covering platforms like GEE and ArcPy, and languages including Python, JavaScript, MATLAB, and R, resulting in 946 complete code samples. Samples already written using the GEE API were directly included in the test set; for the rest, expert judgment was used to determine whether they could be migrated. If deemed migratable, the code was translated into GEE Python API using DeepSeek-V3, and the logic and executability were reviewed by three experts experienced in GEE development. In the end, 88 representative tasks were selected to construct the theme test set. As these samples are complete analysis workflows, they do not require additional function-level parsing and can be directly used as prompts for model generation of test questions (template shown in Figure 12).

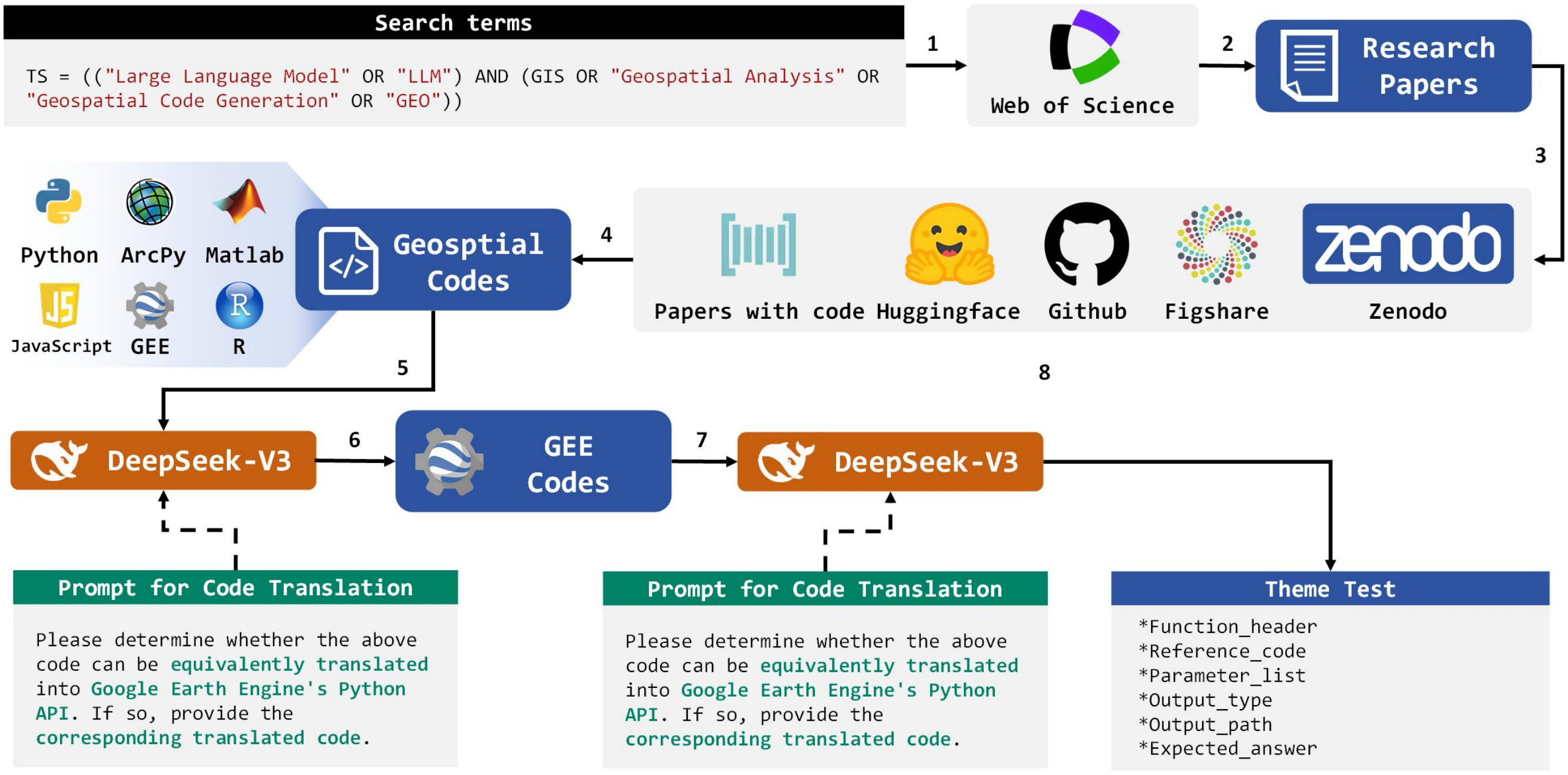

**Figure 11. Theme Test Construction Method.**

```
Promt for Theme Test

## Task Description
You need to generate test code and configuration file entries for given Google Earth Engine (GEE) Python API analysis theme.
The output will have two parts: the standard code and the test cases in the configuration file.

### Input
**Theme**: The required theme of the task

### Output
1. **Standard Code**: Define an analysis function with the theme and returns the result.
The function name should be clear. It should have a docstring that explains the function and its parameters.
2. **Test Cases in Configuration File**: Include multiple test cases, each with parameters, expected answer path, and output type.

### Output Type
1. The output type can be one of the following:
GEE objects:
"ee.Image", "ee.FeatureCollection", "ee.Number", "ee.List", "ee.Dictionary", "ee.Geometry", "ee.Array", "ee.ImageArray"
Python objects:
"str", "int", "float", "bool", "list", "dict"
2. You can use other types if needed, but there MUST be ONE output.

### Expected answer
1. The value of the "expected_answer" field in the configuration file MUST be the path to the file containing the expected output.
2. The file name should be (function name + "_testcase" + testcase_number), file type should be .npy for images and arrays,
 .geojson for geometry or feature objects, .txt for other types.

### Note
1. The tasks are used for automated testing, so they MUST be clear and concise.
2. Do NOT use a whole Image to reduce computation, clip it into a smaller region first.
3. If the output is a GEE object, do NOT perform getInfo() function. Just return the object.
4. Any import statements or initialization statements are NOT needed in the standard code.
5. Do NOT add any explanation or example usages.

### Theme
Here is the theme:
```

**Figure 12. Prompt for theme test construction.**

### 3.7. Construction Results

All test questions were reviewed by experts to ensure the geospatial task definitions were reasonable, problem descriptions were clear, and test case configurations were standardized. For questions with execution errors or logical flaws, experts revised and optimized them based on professional judgment. The manual intervention processes involved in sample selection and code validation in the test set were all completed by three geoscience experts with experience in GEE programming. Their background information and selection criteria are detailed in Table 4.

**Table 4. Background and Selection Criteria of Experts**

| No. | Age | Qualification | Selection Criteria |
|---|---|---|---|
| 1 | 58 | Professor | A senior doctoral advisor with extensive experience in geospatial information systems and service technologies. This expert has published over 120 papers in high-impact international journals, led multiple nationally funded research programs, and contributed significantly to the development of advanced platforms for geospatial computation and analysis. |
| 2 | 35 | Professor | A doctoral supervisor specializing in geospatial modeling, spatiotemporal data mining, and task-oriented analysis. This expert serves on editorial boards of several internationally renowned journals and has engaged in research collaborations with organizations including NASA, NSF, USGS, GEOSS, and Microsoft. Their work bridges both theoretical research and real-world applications, particularly in urban mobility systems and the integration of large-scale geospatial datasets. |
| 3 | 26 | PhD Candidate | An early-career researcher focused on geospatial algorithm design and automated code generation. With a background in computer science (BSc and MSc), this expert has won top honors in national-level GIS programming competitions and has hands-on experience with Earth Engine-based development. Their expertise includes user-oriented evaluation of generated code, emphasizing interpretability, robustness, and ease of execution. |

AutoGEEval++-Bench contains a total of 5,078 unit test cases (including 2,799 general cases and 2,279 boundary cases), 1,199 combination test cases, and 88 theme test cases. In terms of function (operator) count distribution, unit tests focus on a single function call, while combination and theme tests involve the collaborative use of multiple functions. The kernel density estimation (KDE) plot, as shown in Figure 13, presents a complexity-increasing trend: "theme tests (average 16.9 functions) > combination tests (average 5.1 functions) > unit tests (fixed at 1 function)," reflecting a well-defined capability gradient and differentiation.

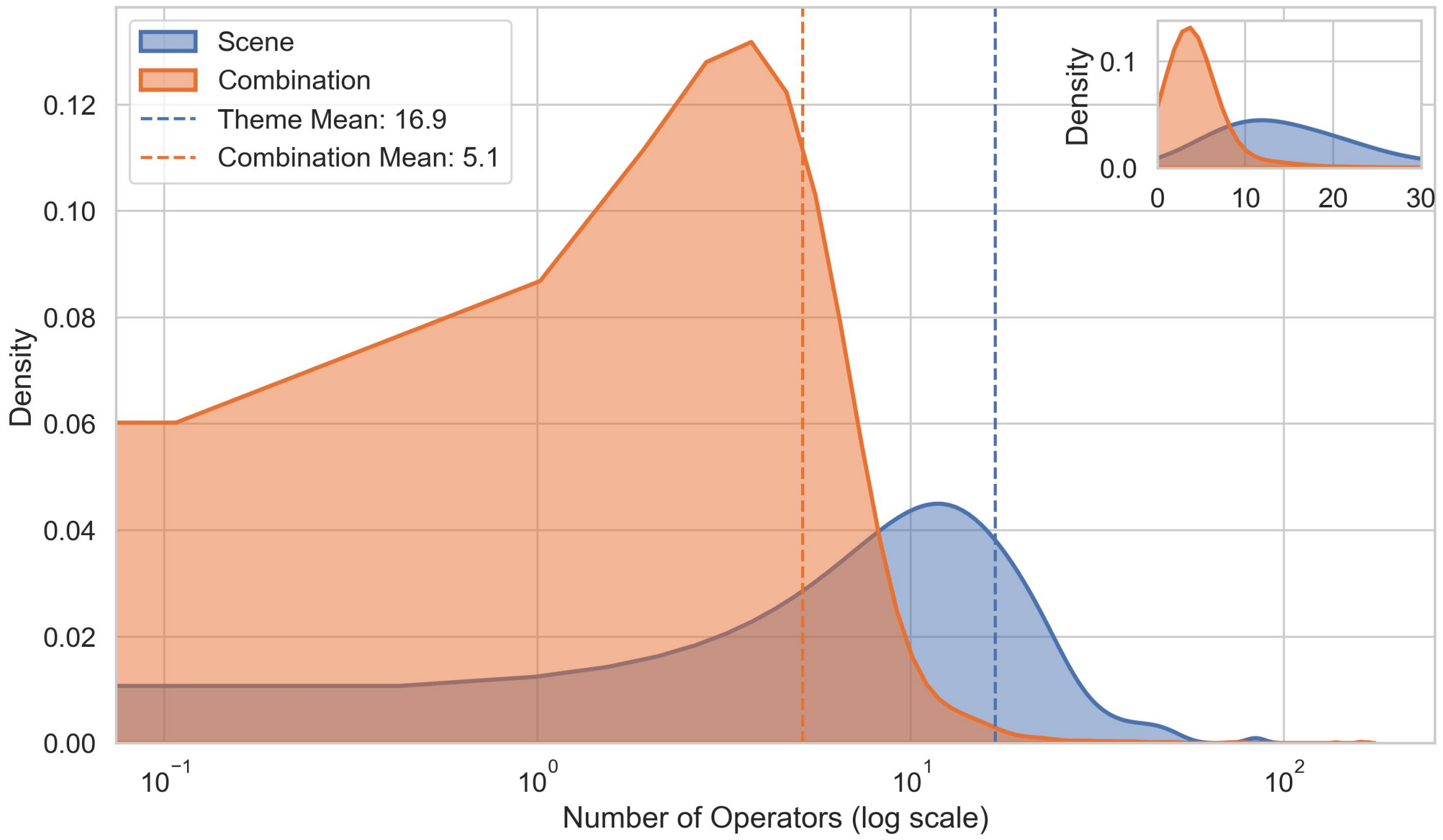

**Figure 13. Operator Count Distribution in Scene-Based vs. Combination-Based Tests.** The x-axis (log scale) denotes the number of operators per combination; the y-axis represents kernel density. Scene-based tests show a broader, right-skewed distribution (mean = 16.9), while combination-based tests cluster around fewer operators (mean = 5.1). The inset zooms into the main density range to emphasize distributional differences.

Figure 14 displays the semantic word clouds of combination tests and theme tests, illustrating their semantic characteristics based on keyword frequency. In the combination tests (Figure 14a), high-frequency terms such as "select", "list", "image", and "map" are mostly related to operator invocation and data structure handling, indicating the coverage of foundational functionalities in GEE programming. In contrast, the theme tests (Figure 14b) feature keywords like "trend", "climate", "population", "composite", and "generation", which reflect more macro-level semantics, focusing on representative themes and application scenarios in remote sensing analysis, and highlighting their problem-driven and integrative analytical characteristics. Overall, the combination tests emphasize function invocation and module organization, whereas the theme tests prioritize contextual integrity and task logic.

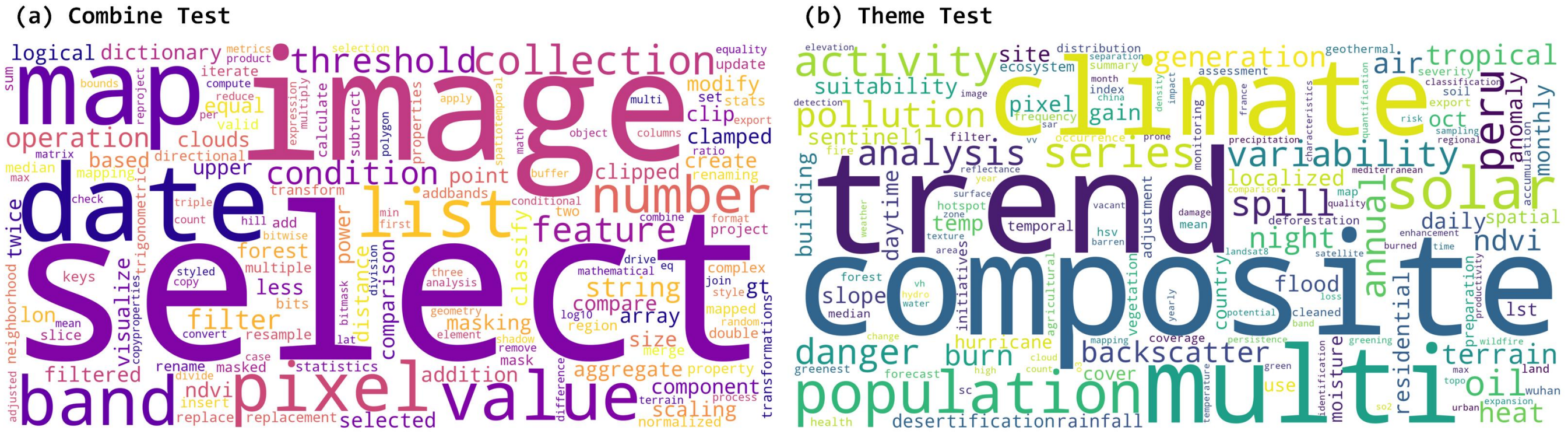

**Figure 14. Semantic word clouds of combination tests (a) and thematic tests (b).**

The test samples constructed in AutoGEEval++-Bench cover all 26 core data types of GEE. The specific number of samples and their proportions are shown in Table 5. From the perspective of data characteristics, these types can be classified into two categories: one is textual formats, such as Dictionary, Array, List, String, Float, and other basic structures; the other is topological formats, including Geometry, Image, FeatureCollection, and GeoJSON, which are spatial expression forms. To demonstrate the coverage capability of the test set across different types, this study presents several representative unit test examples. Among them, Figure 15 shows a sample of the textual format represented by “ee.Array”, and Figure 16 shows a sample of the topological format represented by “ee.Geometry”.

**ee.Array**

**Function_header**

```
def arrayAbsTask(input: ee.Array) -> ee.Array:
    """On an element-wise basis, computes the absolute value of the input."""
```

**Reference_code**

```
def arrayAbsTask(input: ee.Array) -> ee.Array:
    """On an element-wise basis, computes the absolute value of the input."""
    abs_array = input.abs()
    return abs_array
```

**Normal_Parameter_list**

```
arrayAbsTask:
  - params:
      input: !python |
        def get_ee_object():
            import ee
            ee.Initialize()
            array = ee.Array([-1, -2.5, 3, -4.8])
            return array
```

**Boundary_Parameter_list**

```
arrayAbsTask:
  - params:
      input: !python |
        def get_ee_object():
            import ee
            ee.Initialize()
            array = ee.Array([], ee.PixelType.float())
            return array
```

| | Normal | Boundary |
|---|---|---|
| **Output_type** | ee.Array | ee.Array |
| **Output_path** | arrayCatTask_testcase.npy | arrayCatTask_boundarycase.npy |
| **Expected_answer** | [1, 2.5, 3, 4.8] | [] |

**Figure 15. Test example for text-based GEE data type “ee.Array”.** Green highlights denote values from the Normal_Parameter_list; orange highlights indicate those from the Boundary_Parameter_list.

## ee.Geometry

**Function_header**

```python
def geometryConvexHullTask(geometry: ee.Geometry, maxError: float = None, proj: str = None) -> ee.Geometry:
    """Returns the convex hull of the given geometry."""
```

**Reference_code**

```python
def geometryConvexHullTask(geometry: ee.Geometry, maxError: float = None, proj: str = None) -> ee.Geometry:
    """Returns the convex hull of the given geometry."""
    convex_hull = geometry.convexHull(maxError=maxError, proj=proj)
    return convex_hull
```

**Normal_Parameter_list**

```yaml
geometryConvexHullTask:
  - params:
      geometry: !python |
        def get_geometry():
            import ee
            ee.Initialize()
            coords = [[-122.085, 37.422], [-122.085,
37.424], [-122.083, 37.424], [-122.083, 37.422]]
            polygon = ee.Geometry.Polygon(coords)
            return polygon
      maxError: 1.0
      proj: "EPSG:4326"
```

**Boundary_Parameter_list**

```yaml
geometryConvexHullTask:
  - params:
      geometry: !python |
        def get_geometry():
            import ee
            ee.Initialize()
            coords = [[-200.0, 100.0], [200.0, -100.0]]
            invalid_polygon = ee.Geometry.Polygon(coords)
            return invalid_polygon
      maxError: 1.0
      proj: "EPSG:4326"
```

| | Normal | Boundary |
|---|---|---|
| **Output_type** | ee.Geometry | ee.Array |
| **Output_path** | geometryConvexHullTask_testcase.geojson | arrayCatTask_boundarycase.npy |

**Expected_answer** (Normal)

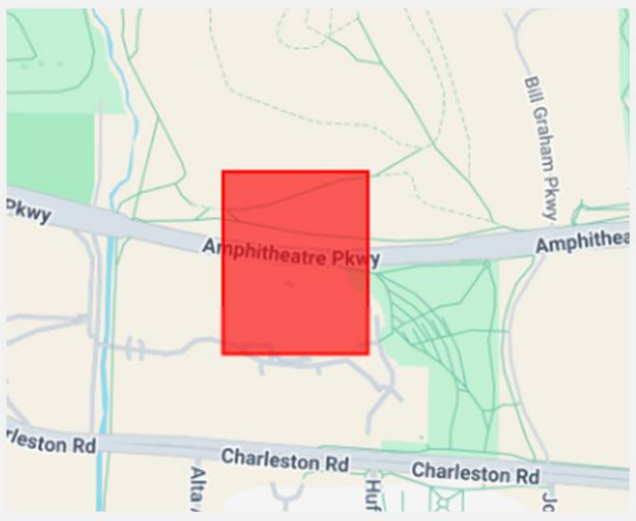

{"geodesic": false, "type": "Polygon", "coordinates": [[[-122.08500000000177, 37.42200000000105], [-122.08299999999807, 37.42200000000105], [-122.08299999999807, 37.42399999999578], [-122.08500000000177, 37.42399999999578], [-122.08500000000177, 37.42200000000105]]]}

**Expected_answer** (Boundary)

EEException: GeometryConstructors.Polygon: LinearRing requires at least 3 points.

**Figure 16. Unit test example for topology-based GEE data type "ee.Geometry".** Green highlights denote values from the Normal_Parameter_list; orange highlights indicate those from the Boundary_Parameter_list.

**Table 5. Distribution of GEE output types in AutoGEEval-Bench.** U.T.Normal refers to general unit tests, U.T.Boundary to boundary unit tests, G.T. to combination tests, and T.T. to thematic tests. Per. indicates the proportion of each type.

| Output_type | U.T.Normal | Per. | U.T.Boundary | Per. | G.T. | Per. | S.T. | Per. |
|---|---|---|---|---|---|---|---|---|
| ee.Array | 254 | 9.07% | 98 | 4.30% | 5 | 0.42% | 2 | 2.27% |
| ee.ArrayImage | 48 | 1.71% | 3 | 0.13% | 1 | 0.08% | 0 | 0.00% |
| ee.Blob | 1 | 0.04% | 0 | 0.00% | 0 | 0.00% | 0 | 0.00% |
| BOOL | 56 | 2.00% | 36 | 1.58% | 4 | 0.33% | 1 | 1.14% |
| ee.Classifier | 28 | 1.00% | 8 | 0.35% | 0 | 0.00% | 0 | 0.00% |
| ee.Clusterer | 16 | 0.57% | 8 | 0.35% | 0 | 0.00% | 0 | 0.00% |
| ee.ConfusionMatrix | 8 | 0.29% | 4 | 0.18% | 0 | 0.00% | 0 | 0.00% |
| ee.Date | 24 | 0.86% | 3 | 0.13% | 0 | 0.00% | 0 | 0.00% |
| ee.DateRange | 11 | 0.39% | 2 | 0.09% | 0 | 0.00% | 0 | 0.00% |
| ee.Dictionary | 118 | 4.22% | 61 | 2.68% | 6 | 0.50% | 10 | 11.36% |
| ee.Element | 16 | 0.57% | 10 | 0.44% | 0 | 0.00% | 0 | 0.00% |
| ee.ErrorMargin | 3 | 0.11% | 1 | 0.04% | 0 | 0.00% | 0 | 0.00% |
| ee.Feature | 34 | 1.21% | 12 | 0.53% | 12 | 1.00% | 0 | 0.00% |
| ee.FeatureCollection | 80 | 2.86% | 54 | 2.37% | 54 | 4.50% | 13 | 14.77% |
| ee.Filter | 90 | 3.22% | 32 | 1.40% | 0 | 0.00% | 0 | 0.00% |
| ee.Geometry | 338 | 12.08% | 113 | 4.96% | 2 | 0.17% | 0 | 0.00% |

| Output_type | U.T.Normal | Per. | U.T.Boundary | Per. | G.T. | Per. | S.T. | Per. |
|---|---|---|---|---|---|---|---|---|
| ee.Image | 420 | 15.01% | 204 | 8.95% | 718 | 59.88% | 44 | 50.00% |
| ee.ImageCollection | 30 | 1.07% | 14 | 0.61% | 334 | 27.86% | 13 | 14.77% |
| ee.Join | 13 | 0.46% | 7 | 0.31% | 0 | 0.00% | 0 | 0.00% |
| ee.Kernel | 56 | 2.00% | 16 | 0.70% | 0 | 0.00% | 0 | 0.00% |
| ee.List | 158 | 5.64% | 65 | 2.85% | 12 | 1.00% | 2 | 2.27% |
| ee.Number | 513 | 18.33% | 133 | 5.84% | 40 | 3.34% | 3 | 3.41% |
| ee.PixelType | 12 | 0.43% | 18 | 0.79% | 0 | 0.00% | 0 | 0.00% |
| ee.Projection | 30 | 1.07% | 9 | 0.39% | 0 | 0.00% | 0 | 0.00% |
| ee.Reducer | 100 | 3.57% | 75 | 3.29% | 0 | 0.00% | 0 | 0.00% |
| ee.String | 342 | 12.22% | 1293 | 56.74% | 11 | 0.92% | 0 | 0.00% |
| **Overall** | **2799** | **100.00%** | **2279** | **100.00%** | **1199** | **100.00%** | **88** | **100.00%** |

## 4. Submission Program and Judge Program

This chapter introduces the process and module design of the Submission program and Judge program in the AutoGEEval++ framework. The former guides LLMs to complete code generation and execution for AutoGEEval++-Bench, while the latter performs comparison and evaluation of the output results.

### 4.1. Submission Program

The overall process of the Submission program is shown in Figure 17 and mainly includes three parts: answering, execution, and result saving. The system sequentially inputs the function declaration (function_header) of the test question into the LLM based on the designed template, guiding it to generate the function implementation. In the execution stage, the system injects actual parameters from the parameters_list into the code and executes the generated logic. The result is saved according to the preset path (output_path) for later use in judgment. The prompt design explicitly requires the model to output only the function body to avoid interference from unstructured text in result parsing.

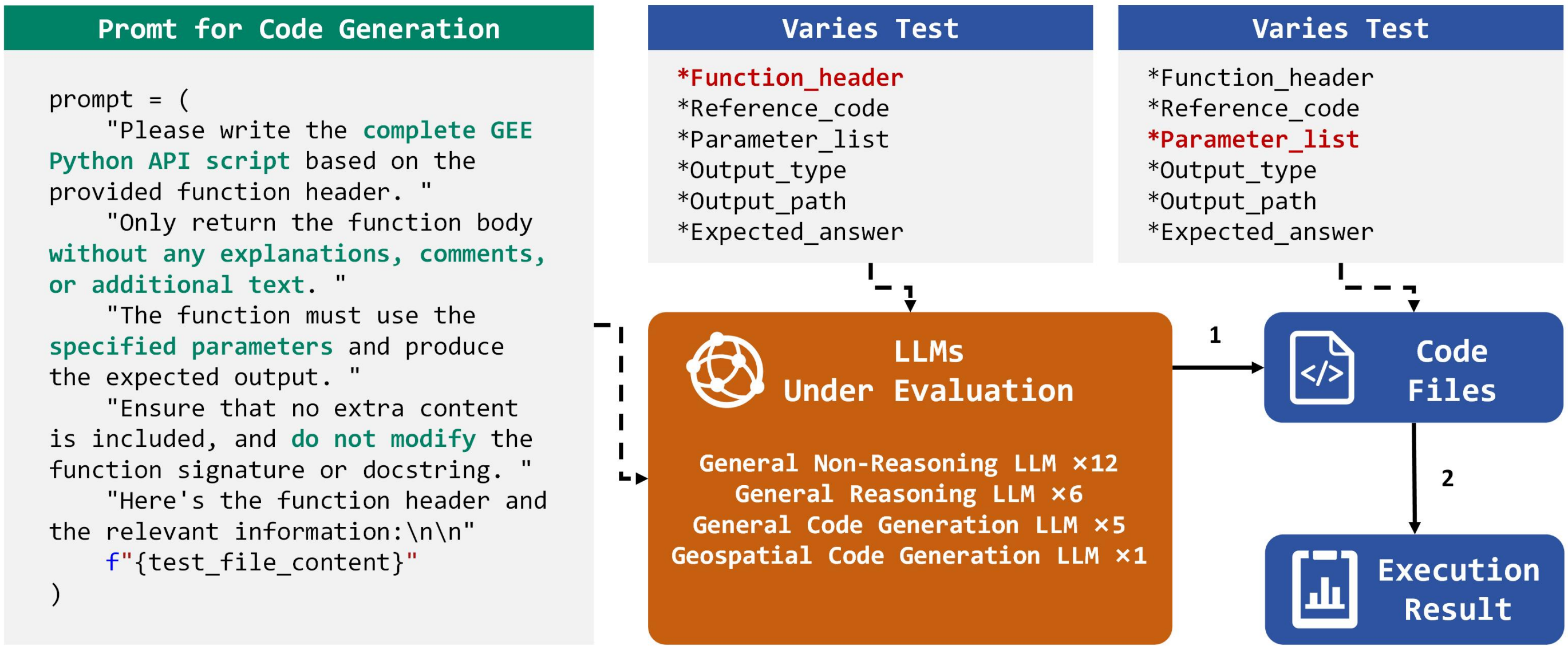

**Figure 17. Submission Program Method.**

### 4.2. Judge Program

The core task of the Judge program is to read the model output stored in output_path and select the corresponding comparison logic based on the output_type to match the result with the expected_answer. The main challenge lies in designing a unified evaluation strategy for multiple data types. GEE provides 26 data types; although the types differ, their numerical representations often overlap. For example, ee.Array, ee.ConfusionMatrix, and ee.ArrayImage are all output in array form; ee.Dictionary, ee.Blob, and ee.Reducer are represented as dictionary structures; ee.Geometry, ee.Feature, and ee.FeatureCollection are commonly expressed in GeoJSON format; and ee.String and ee.Boolean are returned in string form.

To achieve accurate and fair automatic evaluation, the Judge program must classify the results according to their actual representation forms (such as arrays, dictionaries, GeoJSON, strings, etc.) and adopt corresponding comparison strategies. For this purpose, the AutoGEEval framework systematically organizes the output structures of various GEE types and their adapted evaluation methods. The specific mapping relationships are shown in Table 6.

**Table 6. Mapping of GEE Data Types to Evaluation Strategies**

| GEE Data Type | Value Representation | Testing Logic |
|---|---|---|
| ee.Array<br>ee.ConfusionMatrix<br>ee.ArrayImage | **Small-scale array** | Use getInfo to convert to a NumPy array and compare each element with expected_answer. |
| ee.Image<br>ee.ImageCollection | **Large-scale array** | Download the image as a NumPy array and perform pixel-wise comparison; for large images, apply center sampling with a tolerance of 0.001. Merge all images into one and evaluate as a single image. |
| ee.List | **List** | Convert to a Python list via getInfo and compare each element. |
| ee.String<br>ee.BOOL | **String** | Convert to a Python string via getInfo and compare directly. Boolean values are also treated as strings. |
| ee.Number | **Floating-point number** | Convert to a Python float via getInfo and compare with the answer. |
| ee.Dictionary<br>ee.Blob<br>ee.Reducer<br>ee.Filter<br>ee.Classifier<br>ee.Clusterer<br>ee.Pixeltype<br>ee.Join<br>ee.Kernel<br>ee.ErrorMargin<br>ee.Element<br>ee.Projection | **All dictionary keys** | Convert to a Python dictionary via getInfo and compare key-value pairs. |
| ee.Date<br>ee.DateRange | **Dictionary 'value' field** | Use getInfo to obtain a dictionary, extract the 'value' field (timestamp in milliseconds) and compare numerically. |
| ee.Geometry<br>ee.Feature<br>ee.FeatureCollection | **GeoJSON** | Convert to GeoJSON using getInfo and compare geometric consistency with Shapely; for Features, extract geometry before comparison. |

## 5. Experiments

This chapter introduces the model selection criteria, experimental setup, evaluation metrics, and time cost analysis.

### 5.1. Evaluated Models

The evaluated models in this study are LLMs released up to June 2025 that demonstrate both advancement and representativeness. All are open-source or publicly available, with wide accessibility and user impact. The evaluation aims to provide reference for current users who prefer end-to-end and user-friendly solutions. It should be noted that strategies such as prompt engineering, RAG enhancement, or agent-based methods, while capable of optimizing specific tasks, rely on complex design, have unstable performance, and do not alter the base model structure, making it difficult to conduct a unified comparison under this framework. Additionally, these strategies often introduce lengthy prompts, increase token consumption, and affect the fairness and resource efficiency of the evaluation. Therefore, this study focuses on the native performance of base models without incorporating external enhancements. The evaluation covers general-purpose language understanding models, reasoning-enhanced models, general-purpose code generation models, and specialized models for geospatial tasks. Some models include multiple versions with different parameter scales. Treating different parameter versions as separate models, a total of 24 models are included, as detailed in Table 7.

**Table 7. Information of Evaluated LLMs**

| Model Type | Model Name | Developer | Size | Year |
|---|---|---|---|---|
| **General Non-Reasoning** | GPT-4.1 | OpenAI | N/A | 2025 |
| | GPT-4.1-mini | OpenAI | N/A | 2025 |
| | Claude3.7-Sonnet | Anthropic | N/A | 2025 |
| | DeepSeek-V3 | DeepSeek | 671B | 2024 |
| | DeepSeek-V3-0324 | DeepSeek | 685B | 2025 |
| | Gemini2.5-flash-0520 | Google | N/A | 2025 |
| | Qwen-2.5 | Alibaba | 3B, 7B, 32B | 2024 |
| | Qwen-3 | Alibaba | 4B, 8B, 32B | 2025 |
| **General Reasoning** | o4-mini | OpenAI | N/A | 2025 |
| | QwQ-32B | Alibaba | 32B | 2025 |
| | Owen-3-Thinking | Alibaba | 4B, 8B, 32B | 2025 |
| | DeepSeek-R1 | DeepSeek | 671B | 2025 |
| **General Code Generation** | DeepSeek-Coder-V2 | DeepSeek | 16B | 2024 |
| | Qwen2.5-Coder | Alibaba | 3B, 7B, 32B | 2024 |
| | Code-Llama-7B | Meta | 7B | 2023 |
| **Geospatial Code Generation** | GeoCode-GPT-7B | Wuhan University | 7B | 2024 |

### 5.2. Experimental Setup

For hardware and parameter settings, all tests were conducted on a local machine equipped with 32GB RAM and an RTX 4090 GPU. During the model inference phase, open-source models with parameter sizes not exceeding 16B were deployed and run locally using the Ollama tool. For open-source models exceeding 16B and all closed-source models, inference was performed via their official API interfaces, accessing cloud-deployed versions. In terms of parameter configuration, for non-reasoning models, the generation temperature was set to 0.2 to enhance output determinism and stability. For reasoning-enhanced models, temperature parameters were left unset, following established research practices to preserve native reasoning capability. Additionally, the maximum output token length for all models was uniformly set to 4096 to ensure complete output return and avoid truncation due to excessive response length. The runtime durations and task-specific details for each stage are provided in Table 8.

**Table 8. Time Allocation Across Experimental Stages**

| Stages | Time Spent (hours) |
|---|---|
| AutoGEEval++-Bench Construction | 70 |
| Expert Manual Revision | 150 |
| Model Inference and Code Execution | 350 |
| Evaluation of Model Responses | 1120 |
| **Total (All Stages)** | **1690** |

### 5.3. Evaluation Metrics

This study systematically evaluates the performance of LLMs on geospatial code generation tasks from five dimensions: Accuracy metrics, Resource consumption metrics, Operational efficiency metrics, and Error type logs.

#### 5.3.1 Accuracy Metrics

Given the common phenomenon of "hallucination" in LLMs—where the same input may yield inconsistent or unreasonable outputs—single-pass generation cannot fully reflect a model's true capability. Therefore, this study adopts pass@n as the primary accuracy metric, which measures the probability that at least one correct result is produced among *n* independent generations, reflecting the model's output reliability and stability. We evaluate and compare models under n = 1, 3, and 5.

$$pass@n = 1 - \frac{\binom{N-c}{n}}{\binom{N}{n}} \tag{7}$$

Where *N* is the total number of generated samples and *c* is the number of correct results. To distinguish performance across different test types, we denote pass@n for unit tests, combo tests, and theme tests as Unit@n, Combo@n, and Theme@n, respectively.

To further characterize result variability across multiple generations, we introduce the Coefficient of Variation (CV) for stability analysis, defined as the ratio of the standard deviation to the mean:

$$CV = \frac{\sigma}{\mu} \tag{8}$$

A smaller CV indicates more concentrated model outputs and higher stability, suggesting weaker hallucination tendencies. Based on this, we define Stability-Adjusted Accuracy (SA) to jointly evaluate a model's accuracy and stability. This metric uses pass@5 for accuracy and CV for stability:

$$SA = \frac{Pass@5}{1 + CV} \tag{9}$$

#### 5.3.2 Resource Consumption Metrics

These metrics reflect the computational requirements and cost involved in model testing, including four categories:

- **Token Consumption (Tok.):** The average number of tokens consumed per test case. For locally deployed models, it reflects computing resource usage; for commercial API models, it is directly tied to cost. As of April 2025, mainstream model prices per million tokens vary widely: GPT-4 Turbo at $10.00, Claude 3 Opus at $15.00, DeepSeek-Coder at $0.60, and Qwen2-72B at $0.80. Thus, token consumption reflects both resource efficiency and deployment cost-effectiveness.
- **Inference Time (In.T):** The average time (in seconds) to generate each test case, indicating the model's response speed and latency, directly impacting user experience.
- **Code Lines (Co.L):** Number of lines of core executable code in the output, excluding comments and natural language explanations. Compared to tokens, this better reflects the model's actual production capability.
- **Response Lines (Res.L):** Total number of output lines per test case, including code, comments, explanations, and prompts. This comprehensively evaluates the completeness and verbosity of model output. Compared to Co.L, Res.L better reflects differences in human readability and response boundary control: some models prefer concise output, while others include extensive explanations, which may reduce information density and increase token usage.

#### 5.3.3 Operational Efficiency Metrics

These metrics assess model accuracy relative to resource input, reflecting cost-effectiveness and utilization. Based on runtime, token usage, and code structure, three efficiency metrics are defined. To ensure consistency and reduce randomness, all efficiency calculations are based on **SA** derived from 5-shot generation.

- **Inference Efficiency (In.T-E):** Measures the model's ability to achieve accuracy per unit time. It is calculated by dividing SA by the average inference duration (in seconds). This metric reflects the balance between response speed and output quality—shorter inference time and higher accuracy indicate better computational efficiency and interactive performance.

$$In.T - E = \frac{SA}{In.T} \tag{10}$$

- **Token Efficiency (Tok.-E):** Evaluates the accuracy achieved per token consumed during generation. It is calculated by dividing SA by the average token count. This metric emphasizes generation economy and provides a basis for cross-model comparison under cost constraints.

$$Tok. - E = \frac{SA}{Tok.} \tag{11}$$

- **Line Efficiency (Line-E):** Measures the efficiency of generating high-quality executable code, reflecting the model's ability to ensure correctness while controlling structural redundancy and language bloat. This metric enables a more objective evaluation of code compactness and practicality in real-world deployments, particularly in geospatial tasks that require efficient execution.

$$Line - E = \frac{SA}{Res.L/Co.L} \tag{12}$$

#### 5.3.4 Error Type Logs

To support qualitative analysis of model performance, the AutoGEEval framework integrates a runtime error capture mechanism based on the GEE platform. This system can automatically record and classify exception types in model-generated code, providing diagnostic data for subsequent debugging and optimization. The mechanism supports identification of the following five typical error types:

- **Syntax Error:** The code structure violates syntax rules, often caused by missing brackets, spelling mistakes, or misuse of keywords. These are surface-level format issues that prevent interpretation or compilation.
- **Parameter Error:** The model fails to correctly define or invoke preset parameters, resulting in automatic injection failure. Common causes include omitted parameters, incorrect ordering, inconsistent spelling, or failure to explicitly use passed-in variables.
- **Output Type Error:** The code runs successfully but returns a value of an unexpected type, causing downstream processes to break. This stems from inaccurate type modeling or insufficient type validation—for example, returning a list instead of a numerical value, or producing an invalid dictionary structure.
- **Answer Error:** The code appears logically sound and executes successfully with the correct output type, but the result deviates from the standard answer. This reflects issues in algorithm implementation, boundary handling, or reasoning logic, representing a deeper level of logical error.
- **Timeout Error:** Caused by overly complex algorithms, inefficient workflows, or hidden loops, resulting in failure to complete the task within a specified time and subsequent system termination. Alternatively, during interaction with remote services (e.g., GEE), improper request logic or structure (such as illegal nesting or invalid chained calls) can lead to server-side system errors—even if the input parameters are valid—indicating that the code violates execution interface specifications.

## 6. Results

Building upon the evaluation metrics presented in Section 5.3, this chapter provides a systematic analysis and comprehensive interpretation of the performance of various LLMs across four dimensions: accuracy metrics, resource consumption metrics, operational efficiency metrics, and error type logs.

### 6.1. Accuracy with Consideration for Hallucination

The evaluation results for unit tests, compositional tests, and scenario-based tests are presented in Tables 9, 10, and 11, respectively.

**Table 9. Accuracy results for unit tests. Unit@1, Unit@3, and Unit@5 represent percentage scores (%).**

| Category | Model | Unit@1 | Unit@3 | Unit@5 | Unit_CV | Unit_SA |
|---|---|---|---|---|---|---|
| 1 | GPT-4.1 | 69.14 | 78.75 | 80.56 | 0.066 | 75.586 |
| 1 | GPT-4.1-mini | 63.12 | 75.92 | 78.4 | 0.092 | 71.77 |
| 1 | Claude3.7-Sonnet | 67.27 | 73.47 | 74.97 | 0.046 | 71.649 |
| 1 | DeepSeek-V3 | 69.81 | 78 | 81.04 | 0.062 | 76.297 |
| 1 | DeepSeek-V3-0324 | 68.22 | 75.66 | 78.1 | 0.057 | 73.903 |
| 1 | Gemini2.5-Flash-0520 | 70.76 | 77.29 | 79.26 | 0.048 | 75.634 |
| 1 | Qwen-2.5-3B | 33.14 | 45 | 50.26 | 0.167 | 43.057 |
| 1 | Qwen-2.5-7B | 46.32 | 54.71 | 57.76 | 0.091 | 52.924 |
| 1 | Qwen-2.5-32B | 63.33 | 68.06 | 69.99 | 0.042 | 67.19 |
| 1 | Qwen-3-4B | 32.53 | 38.79 | 41.55 | 0.1 | 37.762 |
| 1 | Qwen-3-8B | 47.72 | 53.9 | 56.01 | 0.067 | 52.496 |
| 1 | Qwen-3-32B | 61.93 | 67.01 | 69.18 | 0.046 | 66.137 |
| 2 | o4-mini | 64.81 | 77.96 | 82.02 | 0.098 | 74.697 |
| 2 | QwQ-32B | 56.46 | 72.27 | 76.86 | 0.128 | 68.168 |
| 2 | Qwen-3-4B-Thinking | 28.53 | 46.46 | 54.81 | 0.253 | 43.729 |
| 2 | Qwen-3-8B-Thinking | 35.03 | 54.16 | 61.26 | 0.221 | 50.177 |
| 2 | Qwen-3-32B-Thinking | 52.7 | 71.88 | 77.31 | 0.157 | 66.827 |
| 2 | DeepSeek-R1-0120 | 56.36 | 73.93 | 79.32 | 0.14 | 69.56 |
| 3 | DeepSeek-Coder-V2-16B | 33.95 | 43.3 | 46.99 | 0.132 | 41.492 |
| 3 | Qwen2.5-Coder-3B | 41.18 | 55.22 | 60.63 | 0.157 | 52.421 |
| 3 | Qwen2.5-Coder-7B | 46.1 | 54.47 | 57.35 | 0.091 | 52.584 |
| 3 | Qwen2.5-Coder-32B | 65.16 | 69.83 | 71.58 | 0.039 | 68.87 |
| 3 | Code-Llama-7B | 42.56 | 53.09 | 57.94 | 0.125 | 51.484 |
| 4 | GeoCode-GPT-7B | 58.08 | 72.95 | 79.43 | 0.127 | 70.454 |

**Table 10. Accuracy results for unit tests. Combo@1, Combo@3, and Combo@5 represent percentage scores (%).**

| Category | Model | Combo@1 | Combo@3 | Combo@5 | Combo_CV | Combo_SA |
|---|---|---|---|---|---|---|
| 1 | GPT-4.1 | 81.15 | 87.57 | 89.24 | 0.041 | 85.762 |
| 1 | GPT-4.1-mini | 82.99 | 88.32 | 90.16 | 0.035 | 87.121 |
| 1 | Claude3.7-Sonnet | 85.07 | 87.82 | 88.16 | 0.016 | 86.78 |
| 1 | DeepSeek-V3 | 81.57 | 85.4 | 86.66 | 0.026 | 84.497 |

| Category | Model | Combo@1 | Combo@3 | Combo@5 | Combo_CV | Combo_SA |
|---|---|---|---|---|---|---|
| 1 | DeepSeek-V3-0324 | 80.98 | 86.74 | 88.24 | 0.037 | 85.118 |
| 1 | Gemini2.5-Flash-0520 | 79.32 | 83.65 | 84.9 | 0.029 | 82.512 |
| 1 | Qwen-2.5-3B | 51.96 | 61.8 | 65.64 | 0.096 | 59.872 |
| 1 | Qwen-2.5-7B | 65.8 | 72.64 | 74.98 | 0.055 | 71.088 |
| 1 | Qwen-2.5-32B | 77.06 | 84.99 | 88.32 | 0.057 | 83.59 |
| 1 | Qwen-3-4B | 42.2 | 48.21 | 49.71 | 0.07 | 46.481 |
| 1 | Qwen-3-8B | 53.71 | 58.38 | 60.47 | 0.049 | 57.638 |
| 1 | Qwen-3-32B | 57.55 | 62.55 | 64.14 | 0.046 | 61.336 |
| 2 | o4-mini | 79.82 | 84.65 | 86.41 | 0.033 | 83.624 |
| 2 | QwQ-32B | 79.98 | 86.91 | 88.57 | 0.044 | 84.862 |
| 2 | Qwen-3-4B-Thinking | 54.46 | 66.39 | 70.31 | 0.106 | 63.584 |
| 2 | Qwen-3-8B-Thinking | 57.13 | 69.31 | 74.31 | 0.108 | 67.078 |
| 2 | Qwen-3-32B-Thinking | 73.89 | 84.57 | 87.66 | 0.072 | 81.779 |
| 2 | DeepSeek-R1-0120 | 79.32 | 86.66 | 88.41 | 0.046 | 84.486 |
| 3 | DeepSeek-Coder-V2-16B | 61.38 | 70.39 | 74.23 | 0.078 | 68.831 |
| 3 | Qwen2.5-Coder-3B | 63.97 | 71.48 | 73.73 | 0.06 | 69.567 |
| 3 | Qwen2.5-Coder-7B | 71.56 | 77.56 | 79.48 | 0.044 | 76.111 |
| 3 | Qwen2.5-Coder-32B | 76.23 | 81.57 | 82.9 | 0.036 | 80.025 |
| 3 | Code-Llama-7B | 57.05 | 67.14 | 71.56 | 0.093 | 65.467 |
| 4 | GeoCode-GPT-7B | 73.29 | 85.05 | 89.81 | 0.084 | 82.855 |

**Table 11. Accuracy results for unit tests. Theme@1, Theme@3, and Theme@5 represent percentage scores (%).**

| Category | Model | Theme@1 | Theme@3 | Theme@5 | Theme_CV | Theme_SA |
|---|---|---|---|---|---|---|
| 1 | GPT-4.1 | 35.23 | 51.14 | 57.95 | 0.198 | 48.376 |
| 1 | GPT-4.1-mini | 40.91 | 51.14 | 54.55 | 0.119 | 48.766 |
| 1 | Claude3.7-Sonnet | 47.73 | 60.23 | 61.36 | 0.109 | 55.308 |
| 1 | DeepSeek-V3 | 44.32 | 54.55 | 57.95 | 0.111 | 52.169 |
| 1 | DeepSeek-V3-0324 | 40.91 | 56.82 | 60.23 | 0.16 | 51.926 |
| 1 | Gemini2.5-Flash-0520 | 34.09 | 46.59 | 50 | 0.157 | 43.215 |
| 1 | Qwen-2.5-3B | 2.27 | 7.95 | 9.09 | 0.463 | 6.212 |
| 1 | Qwen-2.5-7B | 17.05 | 23.86 | 26.14 | 0.173 | 22.289 |
| 1 | Qwen-2.5-32B | 29.55 | 43.18 | 46.59 | 0.185 | 39.313 |
| 1 | Qwen-3-4B | 9.09 | 12.5 | 12.5 | 0.142 | 10.951 |
| 1 | Qwen-3-8B | 11.36 | 15.91 | 17.05 | 0.166 | 14.618 |
| 1 | Qwen-3-32B | 30.68 | 38.64 | 39.77 | 0.111 | 35.789 |
| 2 | o4-mini | 43.18 | 59.09 | 64.77 | 0.164 | 55.639 |
| 2 | QwQ-32B | 42.05 | 51.14 | 54.55 | 0.107 | 49.272 |
| 2 | Qwen-3-4B-Thinking | 7.95 | 14.77 | 15.91 | 0.273 | 12.499 |
| 2 | Qwen-3-8B-Thinking | 15.91 | 23.86 | 25 | 0.187 | 21.057 |
| 2 | Qwen-3-32B-Thinking | 25 | 47.73 | 52.27 | 0.286 | 40.635 |
| 2 | DeepSeek-R1-0120 | 37.5 | 50 | 57.95 | 0.174 | 49.378 |
| 3 | DeepSeek-Coder-V2-16B | 11.36 | 14.77 | 20.45 | 0.242 | 16.472 |

| Category | Model | Theme@1 | Theme@3 | Theme@5 | Theme_CV | Theme_SA |
|---|---|---|---|---|---|---|
| 3 | Qwen2.5-Coder-3B | 10.23 | 15.91 | 18.18 | 0.226 | 14.825 |
| 3 | Qwen2.5-Coder-7B | 18.18 | 23.86 | 32.95 | 0.243 | 26.501 |
| 3 | Qwen2.5-Coder-32B | 30.68 | 43.18 | 45.45 | 0.163 | 39.07 |
| 3 | Code-Llama-7B | 6.82 | 11.36 | 13.64 | 0.267 | 10.763 |
| 4 | GeoCode-GPT-7B | 24.55 | 26.82 | 30.23 | 0.086 | 27.841 |

Figure 18 illustrates the accuracy and multi-round generation gain trends of major LLMs across the three task types (Unit tests, Combo tests, and Theme tests) under the AutoGEEval++ framework. Overall, the models exhibit a consistent performance hierarchy of "Combo > Unit > Theme", which can be attributed to the following factors: First, imbalanced training data distribution. Current large models are primarily trained on sources such as GitHub, API documentation, and Q&A forums. Function composition patterns (e.g., image selection + band processing + export) frequently appear in training corpora, facilitating *"memory matching"* in combo tests. In contrast, coverage of official GEE function documentation is relatively limited, while theme tests involve multimodal and semantically structured tasks that are much scarcer in the training data. Second, increasing task complexity. Combo tests feature clear structure and high error tolerance; unit tests demand precise understanding of function semantics and parameter constraints; theme tests span entire workflows, requiring global planning and cross-module orchestration, which significantly increases cognitive load. Third, current training paradigms are biased toward function and composition generation, lacking systematic modeling of task comprehension, semantic alignment, and module integration. As a result, models are prone to structural omissions or hallucinations in theme tests. Additionally, the line chart shows that the accuracy improvement from @3 over @1 is significantly greater than from @5 over @3, indicating that multi-round generation can alleviate initial hallucination, but with diminishing marginal returns, suggesting an upper bound to overall improvement.

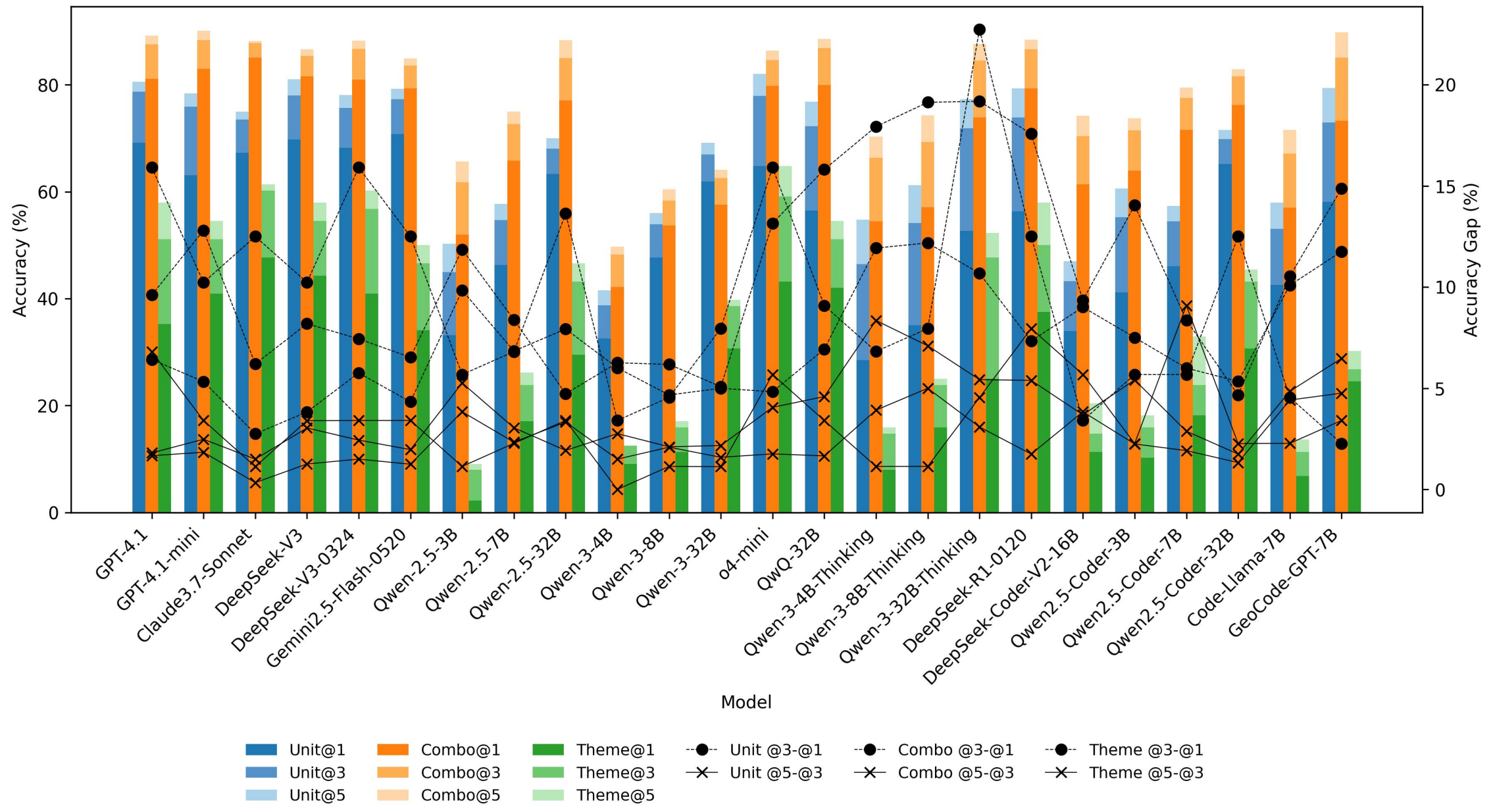

**Figure 18. Stacked Accuracy Distribution and Improvement Trends Across Three Test Sets for Different Models.** Stacked bars and line plots show model accuracies on Unit (blue), Combo (orange), and Theme (green) tests at @1, @3, and @5. X-axis: models; left Y-axis: accuracy (%); right Y-axis: improvement (%). Dashed lines (○) indicate gains from @1→@3; solid lines (×) from @3→@5.

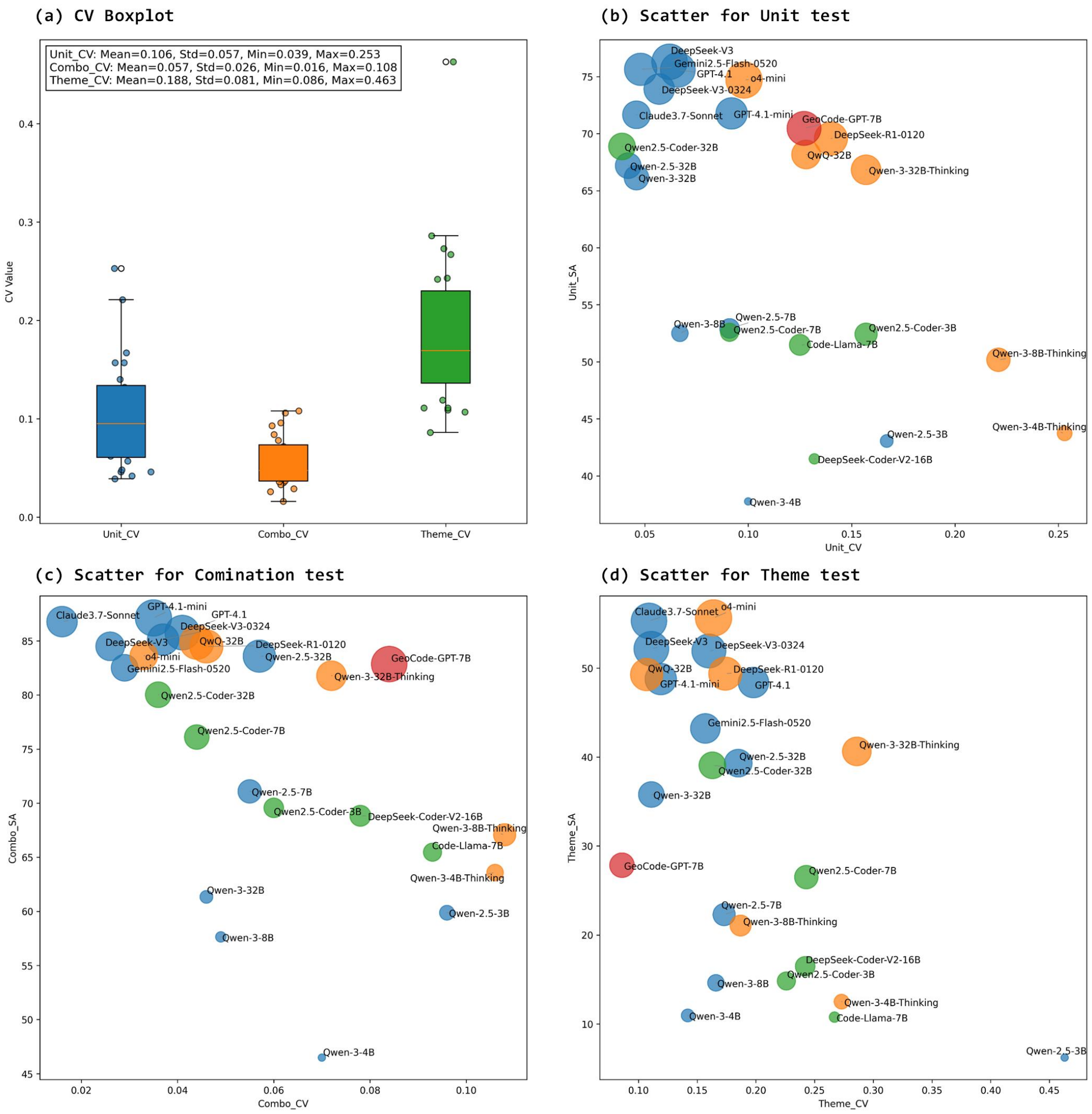

**Figure 19. Visualization of CV Distribution and Performance Correlation across Test Types.** Subfigure (a) shows the CV boxplot for Unit, Combo, and Theme tests; subfigures (b)–(d) present scatter plots of CV versus overall score. Bubble size reflects performance rank; color indicates model category.

Figure 18 shows the distribution of Coefficient of Variation (CV) for different test types and its correlation with performance metrics. As shown in Figure 19a, CV is lowest for combination tests, higher for unit tests, and highest for thematic tests, indicating that combination test outputs are more stable while thematic tests are more prone to structural hallucinations. Figures 19b–19d demonstrate that models with lower CV, higher SA, and higher Pass@5 scores tend to perform better, with data points clustering in the upper-left region. Semantically, SA can be viewed as accuracy under hallucination control. The ranking of SA across test types is summarized in Table 12.

**Table 12. Stability-aware Accuracy (SA) Rankings of LLMs Across Different Test Types.** Blue indicates general non-reasoning models, orange indicates general reasoning models, green indicates general code generation models, and red indicates geospatial code generation models.

| Rank | Unit_SA | Combo_SA | Theme_SA |
|---|---|---|---|
| 1 | DeepSeek-V3 | GPT-4.1-mini | o4-mini |
| 2 | Gemini2.5-Flash-0520 | Claude3.7-Sonnet | Claude3.7-Sonnet |
| 3 | GPT-4.1 | GPT-4.1 | DeepSeek-V3 |
| 4 | o4-mini | DeepSeek-V3-0324 | DeepSeek-V3-0324 |
| 5 | DeepSeek-V3-0324 | QwQ-32B | DeepSeek-R1-0120 |
| 6 | GPT-4.1-mini | DeepSeek-V3 | QwQ-32B |

| Rank | Unit_SA | Combo_SA | Theme_SA |
|---|---|---|---|
| 7 | Claude3.7-Sonnet | DeepSeek-R1-0120 | GPT-4.1-mini |
| 8 | GeoCode-GPT-7B | o4-mini | GPT-4.1 |
| 9 | DeepSeek-R1-0120 | Qwen-2.5-32B | Gemini2.5-Flash-0520 |
| 10 | Qwen2.5-Coder-32B | GeoCode-GPT-7B | Qwen-3-32B-Thinking |
| 11 | QwQ-32B | Gemini2.5-Flash-0520 | Qwen-2.5-32B |
| 12 | Qwen-2.5-32B | Qwen-3-32B-Thinking | Qwen2.5-Coder-32B |
| 13 | Qwen-3-32B-Thinking | Qwen2.5-Coder-32B | Qwen-3-32B |
| 14 | Qwen-3-32B | Qwen2.5-Coder-7B | GeoCode-GPT-7B |
| 15 | Qwen-2.5-7B | Qwen-2.5-7B | Qwen2.5-Coder-7B |
| 16 | Qwen2.5-Coder-7B | Qwen2.5-Coder-3B | Qwen-2.5-7B |
| 17 | Qwen-3-8B | DeepSeek-Coder-V2-16B | Qwen-3-8B-Thinking |
| 18 | Qwen2.5-Coder-3B | Qwen-3-8B-Thinking | DeepSeek-Coder-V2-16B |
| 19 | Code-Llama-7B | Code-Llama-7B | Qwen2.5-Coder-3B |
| 20 | Qwen-3-8B-Thinking | Qwen-3-4B-Thinking | Qwen-3-8B |
| 21 | Qwen-3-4B-Thinking | Qwen-3-32B | Qwen-3-4B-Thinking |
| 22 | Qwen-2.5-3B | Qwen-2.5-3B | Qwen-3-4B |
| 23 | DeepSeek-Coder-V2-16B | Qwen-3-8B | Code-Llama-7B |
| 24 | Qwen-3-4B | Qwen-3-4B | Qwen-2.5-3B |

In terms of model types, general-purpose non-reasoning models (blue) consistently rank higher across all three test categories and show strong stability. General-purpose reasoning models (orange) follow, demonstrating an advantage in tasks with higher semantic complexity. General-purpose code generation models (green) generally perform in the mid-to-lower range, while geospatial code models (red) are positioned in the middle. In more structurally complex and context-dependent thematic tests, general-purpose reasoning models perform particularly well—lightweight model "o4-mini" and mid-size model "DeepSeek-R1-0120" rank 1st and 5th respectively, significantly outperforming their rankings in other test types, indicating strong reasoning and task organization capabilities.

Regarding model families, the GPT series shows overall consistency: "GPT-4.1" and "GPT-4.1-mini" have relatively stable rankings across tasks (3/3/8 and 6/1/7 respectively), with the mini version outperforming the full version in combination and thematic tests, possibly due to enhanced generalization from distillation and compression. DeepSeek general-purpose models (V3 and R1) consistently rank in the top nine, balancing performance and cost. However, the code-specific model "DeepSeek-Coder-V2-16B" performs poorly (23/17/18), indicating room for improvement in multi-layer, cross-function tasks. "Claude3.7-Sonnet" ranks 2nd in combination and thematic tests, but 7th in unit tests, suggesting it excels at structural generation but is less precise in atomic function calls. "Gemini2.5-Flash-0520" ranks 2nd in unit tests but drops to 11th and 9th in combination and thematic tests respectively, reflecting relatively weaker chain-of-thought reasoning and task planning, making it more suitable for basic function-level tasks.

As for parameter size and version iterations, performance generally improves with model scale within the same series. For instance, "Qwen-2.5-32B" outperforms its 7B and 3B counterparts. However, version upgrades do not guarantee performance improvement—the Qwen-3 series underperforms relative to "Qwen-2.5" of the same size, suggesting possible degradation in training strategies. Mid-size domain-specific model GeoCode-GPT-7B outperforms several general-purpose models (e.g., "Qwen-2.5-7B", "Code-Llama-7B", and "Qwen-3-8B-Thinking"), and even surpasses "DeepSeek-R1-0120" and "Qwen2.5-Coder-32B" in certain tasks, highlighting the effectiveness of domain fine-tuning in geospatial generation tasks. Lightweight models "o4-mini" and "GPT-4.1-mini" repeatedly rank in the top three for combination and thematic tests, demonstrating outstanding parameter compression and stability, making them suitable for high-accuracy, low-resource applications.

### 6.2. Resource Consumption and Operational Efficiency

The resource consumption and operational efficiency results of each model across the three types of test tasks are presented in Tables 13 through 15, respectively.

**Table 13. Resource Consumption and Runtime Efficiency for Unit Tests.**

| Category | Model | Tok. | In.T | Co.L | Res.L | Tok.-E | In.T-E | Line-E |
|---|---|---|---|---|---|---|---|---|
| 1 | GPT-4.1 | 314.94 | 6.28 | 14.79 | 17.79 | 0.240 | 12.036 | 62.840 |
| 1 | GPT-4.1-mini | 308.99 | 6.75 | 13.85 | 16.86 | 0.232 | 10.633 | 58.957 |
| 1 | Claude3.7-Sonnet | 451.40 | 5.35 | 20.64 | 27.62 | 0.159 | 13.392 | 53.542 |
| 1 | DeepSeek-V3 | 257.51 | 3.93 | 7.23 | 10.24 | 0.296 | 19.414 | 53.870 |
| 1 | DeepSeek-V3-0324 | 271.40 | 5.10 | 9.34 | 12.35 | 0.272 | 14.491 | 55.891 |
| 1 | Gemini2.5-Flash-0520 | 410.19 | 5.11 | 20.34 | 23.36 | 0.184 | 14.801 | 65.856 |
| 1 | Qwen-2.5-3B | 266.98 | 1.16 | 9.68 | 12.66 | 0.161 | 37.118 | 32.922 |
| 1 | Qwen-2.5-7B | 287.81 | 1.75 | 9.61 | 15.86 | 0.184 | 30.242 | 32.068 |
| 1 | Qwen-2.5-32B | 273.69 | 4.14 | 9.63 | 12.63 | 0.245 | 16.229 | 51.230 |
| 1 | Qwen-3-4B | 314.56 | 2.13 | 10.80 | 13.83 | 0.120 | 17.729 | 29.489 |
| 1 | Qwen-3-8B | 310.76 | 2.72 | 9.77 | 12.86 | 0.169 | 19.300 | 39.882 |
| 1 | Qwen-3-32B | 288.04 | 4.08 | 10.03 | 17.04 | 0.230 | 16.210 | 38.929 |
| 2 | o4-mini | 1705.47 | 21.10 | 13.92 | 16.94 | 0.044 | 3.540 | 61.380 |
| 2 | QwQ-32B | 2800.08 | 82.88 | 16.09 | 197.20 | 0.024 | 0.822 | 5.562 |
| 2 | Qwen-3-4B-Thinking | 2578.33 | 53.71 | 43.81 | 123.15 | 0.017 | 0.814 | 15.556 |
| 2 | Qwen-3-8B-Thinking | 2729.17 | 67.30 | 46.57 | 133.95 | 0.018 | 0.746 | 17.445 |
| 2 | Qwen-3-32B-Thinking | 2601.18 | 86.81 | 13.58 | 160.24 | 0.026 | 0.770 | 5.663 |
| 2 | DeepSeek-R1-0120 | 3489.02 | 120.93 | 9.05 | 208.62 | 0.020 | 0.575 | 3.018 |
| 3 | DeepSeek-Coder-V2-16B | 350.79 | 6.33 | 12.88 | 16.33 | 0.118 | 6.555 | 32.726 |
| 3 | Qwen2.5-Coder-3B | 349.46 | 2.56 | 12.16 | 23.66 | 0.150 | 20.477 | 26.942 |
| 3 | Qwen2.5-Coder-7B | 314.85 | 2.31 | 10.75 | 18.32 | 0.167 | 22.764 | 30.856 |
| 3 | Qwen2.5-Coder-32B | 295.73 | 6.98 | 11.42 | 14.44 | 0.233 | 9.867 | 54.466 |
| 3 | Code-Llama-7B | 334.48 | 1.79 | 9.03 | 12.96 | 0.154 | 28.762 | 35.872 |
| 4 | GeoCode-GPT-7B | 320.58 | 2.49 | 13.33 | 18.05 | 0.220 | 28.295 | 52.031 |

**Table 14. Resource Consumption and Runtime Efficiency for Combination Tests.**

| Category | Model | Tok. | In.T | Co.L | Res.L | Tok.-E | In.T-E | Line-E |
|---|---|---|---|---|---|---|---|---|
| 1 | GPT-4.1 | 730.81 | 4.21 | 9.23 | 11.24 | 0.093 | 20.371 | 70.426 |
| 1 | GPT-4.1-mini | 919.41 | 7.87 | 8.05 | 26.08 | 0.095 | 11.070 | 26.891 |
| 1 | Claude3.7-Sonnet | 882.31 | 7.53 | 14.81 | 17.14 | 0.098 | 11.525 | 74.983 |
| 1 | DeepSeek-V3 | 729.31 | 5.19 | 9.55 | 11.55 | 0.116 | 16.281 | 69.865 |
| 1 | DeepSeek-V3-0324 | 726.63 | 5.17 | 9.50 | 11.52 | 0.117 | 16.464 | 70.193 |
| 1 | Gemini2.5-Flash-0520 | 922.37 | 3.76 | 12.55 | 22.34 | 0.089 | 21.945 | 46.353 |
| 1 | Qwen-2.5-3B | 777.97 | 2.26 | 12.24 | 14.84 | 0.077 | 26.492 | 49.382 |
| 1 | Qwen-2.5-7B | 754.95 | 2.23 | 11.20 | 13.79 | 0.094 | 31.878 | 57.736 |
| 1 | Qwen-2.5-32B | 746.1 | 10.28 | 11.48 | 14.30 | 0.112 | 8.131 | 67.106 |
| 1 | Qwen-3-4B | 878.75 | 4.51 | 11.39 | 21.12 | 0.053 | 10.306 | 25.067 |
| 1 | Qwen-3-8B | 954.96 | 7.47 | 6.56 | 28.07 | 0.060 | 7.716 | 13.470 |
| 1 | Qwen-3-32B | 735.39 | 3.47 | 9.02 | 15.11 | 0.083 | 17.676 | 36.615 |
| 2 | o4-mini | 1340.57 | 8.99 | 6.91 | 9.75 | 0.062 | 9.302 | 59.266 |
| 2 | QwQ-32B | 2043.27 | 52.81 | 9.45 | 119.75 | 0.042 | 1.607 | 6.697 |
| 2 | Qwen-3-4B-Thinking | 2455.26 | 33.05 | 20.86 | 120.64 | 0.026 | 1.924 | 10.994 |
| 2 | Qwen-3-8B-Thinking | 2649.54 | 36.73 | 11.54 | 131.10 | 0.025 | 1.826 | 5.905 |
| 2 | Qwen-3-32B-Thinking | 1902.56 | 43.09 | 7.30 | 79.31 | 0.043 | 1.898 | 7.527 |
| 2 | DeepSeek-R1-0120 | 2227.44 | 55.91 | 6.06 | 94.59 | 0.038 | 1.511 | 5.413 |

| Category | Model | Tok. | In.T | Co.L | Res.L | Tok.-E | In.T-E | Line-E |
|---|---|---|---|---|---|---|---|---|
| 3 | DeepSeek-Coder-V2-16B | 1064.84 | 11.96 | 12.21 | 30.66 | 0.065 | 5.755 | 27.411 |
| 3 | Qwen2.5-Coder-3B | 1133.21 | 9.32 | 15.85 | 47.01 | 0.061 | 7.464 | 23.455 |
| 3 | Qwen2.5-Coder-7B | 749.98 | 2.14 | 9.60 | 12.76 | 0.101 | 35.566 | 57.262 |
| 3 | Qwen2.5-Coder-32B | 741.94 | 11.58 | 11.53 | 13.53 | 0.108 | 6.911 | 68.196 |
| 3 | Code-Llama-7B | 1185.59 | 8.81 | 13.29 | 34.83 | 0.055 | 7.431 | 24.980 |
| 4 | GeoCode-GPT-7B | 769.91 | 2.66 | 12.38 | 15.08 | 0.108 | 31.148 | 68.020 |

**Table 15. Resource Consumption and Runtime Efficiency for Theme Tests.**

| Category | Model | Tok. | In.T | Co.L | Res.L | Tok.-E | In.T-E | Line-E |
|---|---|---|---|---|---|---|---|---|
| 1 | GPT-4.1 | 849.25 | 12.18 | 33.51 | 35.62 | 0.057 | 3.972 | 45.510 |
| 1 | GPT-4.1-mini | 1109.28 | 15.78 | 34.34 | 58.28 | 0.044 | 3.090 | 28.734 |
| 1 | Claude3.7-Sonnet | 1064.98 | 11.45 | 38.10 | 40.13 | 0.052 | 4.830 | 52.510 |
| 1 | DeepSeek-V3 | 744.74 | 9.17 | 21.33 | 23.33 | 0.070 | 5.689 | 47.697 |
| 1 | DeepSeek-V3-0324 | 768.5 | 8.93 | 25.10 | 27.34 | 0.068 | 5.815 | 47.672 |
| 1 | Gemini2.5-Flash-0520 | 1279.08 | 7.07 | 39.71 | 60.79 | 0.034 | 6.112 | 28.229 |
| 1 | Qwen-2.5-3B | 828.28 | 5.22 | 27.13 | 29.82 | 0.007 | 1.190 | 5.652 |
| 1 | Qwen-2.5-7B | 809.1 | 5.52 | 26.02 | 30.13 | 0.028 | 4.038 | 19.249 |
| 1 | Qwen-2.5-32B | 771.55 | 17.19 | 24.88 | 26.90 | 0.051 | 2.287 | 36.361 |
| 1 | Qwen-3-4B | 1160.46 | 12.21 | 33.50 | 58.76 | 0.009 | 0.897 | 6.243 |
| 1 | Qwen-3-8B | 1203 | 16.11 | 28.29 | 58.53 | 0.012 | 0.907 | 7.065 |
| 1 | Qwen-3-32B | 780.85 | 7.81 | 24.91 | 32.00 | 0.046 | 4.582 | 27.859 |
| 2 | o4-mini | 2145.06 | 24.13 | 24.31 | 29.25 | 0.026 | 2.306 | 46.242 |
| 2 | QwQ-32B | 3440.17 | 78.47 | 35.46 | 274.27 | 0.014 | 0.628 | 6.370 |
| 2 | Qwen-3-4B-Thinking | 4051.27 | 80.17 | 124.11 | 203.93 | 0.003 | 0.156 | 7.607 |
| 2 | Qwen-3-8B-Thinking | 4023.85 | 92.16 | 120.60 | 201.74 | 0.005 | 0.228 | 12.588 |
| 2 | Qwen-3-32B-Thinking | 3282.77 | 88.36 | 29.66 | 193.56 | 0.012 | 0.460 | 6.227 |
| 2 | DeepSeek-R1-0120 | 4119.38 | 128.52 | 20.90 | 235.63 | 0.012 | 0.384 | 4.380 |
| 3 | DeepSeek-Coder-V2-16B | 1139.58 | 17.68 | 28.12 | 46.25 | 0.014 | 0.932 | 10.015 |
| 3 | Qwen2.5-Coder-3B | 1166.05 | 11.80 | 31.41 | 61.29 | 0.013 | 1.256 | 7.598 |
| 3 | Qwen2.5-Coder-7B | 815.92 | 5.70 | 25.32 | 30.33 | 0.032 | 4.649 | 22.123 |
| 3 | Qwen2.5-Coder-32B | 777.41 | 17.58 | 26.55 | 28.55 | 0.050 | 2.222 | 36.333 |
| 3 | Code-Llama-7B | 1116.18 | 8.78 | 23.13 | 35.17 | 0.010 | 1.226 | 7.078 |
| 4 | GeoCode-GPT-7B | 839.02 | 6.47 | 26.55 | 31.75 | 0.033 | 4.303 | 23.281 |

Token consumption across the three test types for each model is shown in Figure 20. Overall, general-purpose reasoning models exhibit significantly higher token usage—often 4 to 8 times greater than other model types—indicating high resource overhead when handling complex reasoning tasks. Most models follow a consumption pattern of "thematic > combination > unit" tests, reflecting increased input information requirements with rising task semantic complexity. However, exceptions are observed among general-purpose reasoning models, where token consumption in unit tests exceeds that of combination tests. This may result from the models introducing more context completion or chain-of-thought reasoning even in simplified inputs, leading to redundant generation and increased consumption.

Notably, although categorized as a reasoning model, o4-mini demonstrates significantly lower token usage, suggesting effective optimization in reasoning strategy, context compression, or output control. In the Qwen series, smaller parameter models sometimes consume more tokens than larger ones; for instance, Qwen-3-4B-Thinking and Qwen-3-8B-Thinking both exceed Qwen-3-32B-Thinking across all test types. Similar trends are observed across several Qwen-2.5 and Qwen-3 models, possibly due to inefficiencies in smaller models such as redundant content, weaker context control, or suboptimal compression strategies, resulting in lower token usage efficiency.

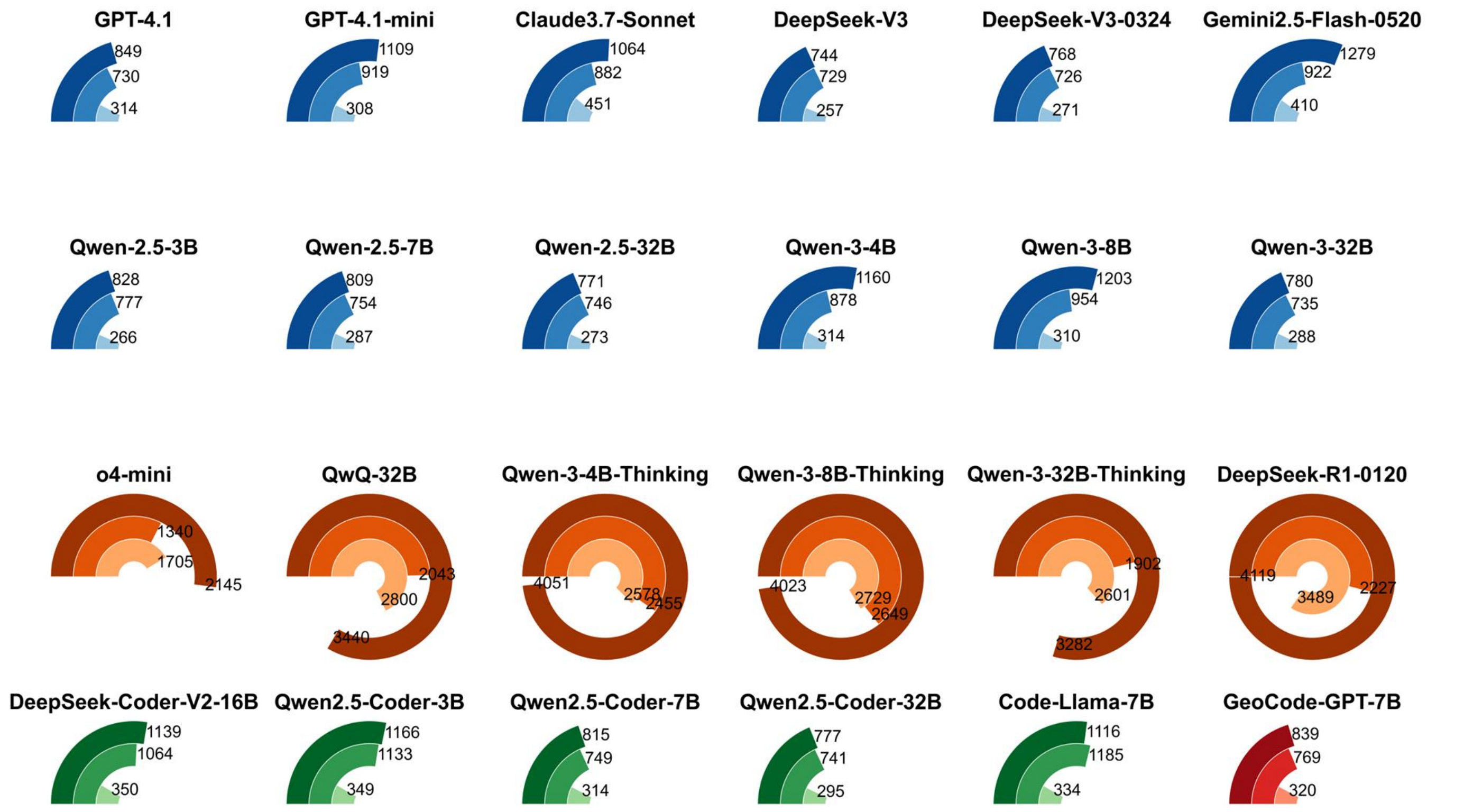

**Figure 20. Token Consumption of Models Across Three Testing Tasks.** Colors represent model types: blue for general non-reasoning, orange for general reasoning, green for general code generation, and red for geospatial code generation. The inner, middle, and outer rings correspond to unit, integration, and thematic tests, respectively. Ring length indicates Token usage.

Figure 21 shows the inference time consumption of each model across the three task types. General-purpose reasoning models (orange) exhibit significantly longer inference times compared to other categories, while general non-reasoning models, general code models, and geospatial models show relatively small differences. Although o4-mini belongs to the reasoning model category, it records the lowest inference time, close to other categories, which may benefit from optimization in reasoning strategy or architecture.

In terms of parameter size comparison, 7B/8B models generally offer better inference efficiency. For example, GeoCode-GPT has the shortest inference time at this scale. Even within the same model series, such as Qwen or LLaMA, the inference times of 7B/8B models are lower than those of the 3B and 32B versions. Small models, due to limited expressive capacity, often compensate by generating redundant information, which in turn prolongs inference time. Large models, though strong in reasoning, have high per-step computation cost due to the large number of parameters, resulting in increased latency. The 7B/8B models strike a good balance between reasoning ability and computational efficiency, making them the mainstream choice for real-world deployment. This also explains why many domain-specific models tend to adopt this scale.

Figure 22 illustrates the relationship between core code lines (Co.L) and total response lines (Res.L) across the three tasks for each model. Since Res.L includes reasoning and explanation, it is generally greater than Co.L, so all data points lie above the gray diagonal line. Points closer to the diagonal indicate that the generated content is closer to core code with less redundancy; those closer to the origin suggest higher expressive efficiency while maintaining completeness. Overall, general-purpose reasoning models show lower expressive compactness, with points concentrated in the upper left, indicating more redundancy. However, o4-mini demonstrates the highest expressive efficiency in combination tests, showing that its lightweight design optimizes output precision. Both versions of DeepSeek-V3 consistently rank among the top four across all tasks, reflecting high consistency and compact generation capability. The Qwen-2.5 series shows higher compactness in unit tests, while the Qwen-2.5-Coder series performs well in thematic tests, indicating task-specific optimization.

Figure 23 presents the accuracy ranking adjusted for efficiency and stability. The two versions of DeepSeek-V3 rank first and second, followed by GPT-4.1. Despite having only 7B parameters, GeoCode-GPT-7B ranks fourth, verifying the effectiveness of its fine-tuning strategy in vertical tasks. General non-reasoning models like Claude 3.7-Sonnet and Gemini 2.5-Flash-0520 also rank highly. Although various versions of Qwen-2.5, Qwen-2.5-Coder, and Qwen-3 exhibit only moderate accuracy, they achieve strong overall performance due to high inference efficiency. In contrast, reasoning models and small-parameter models generally rank lower.

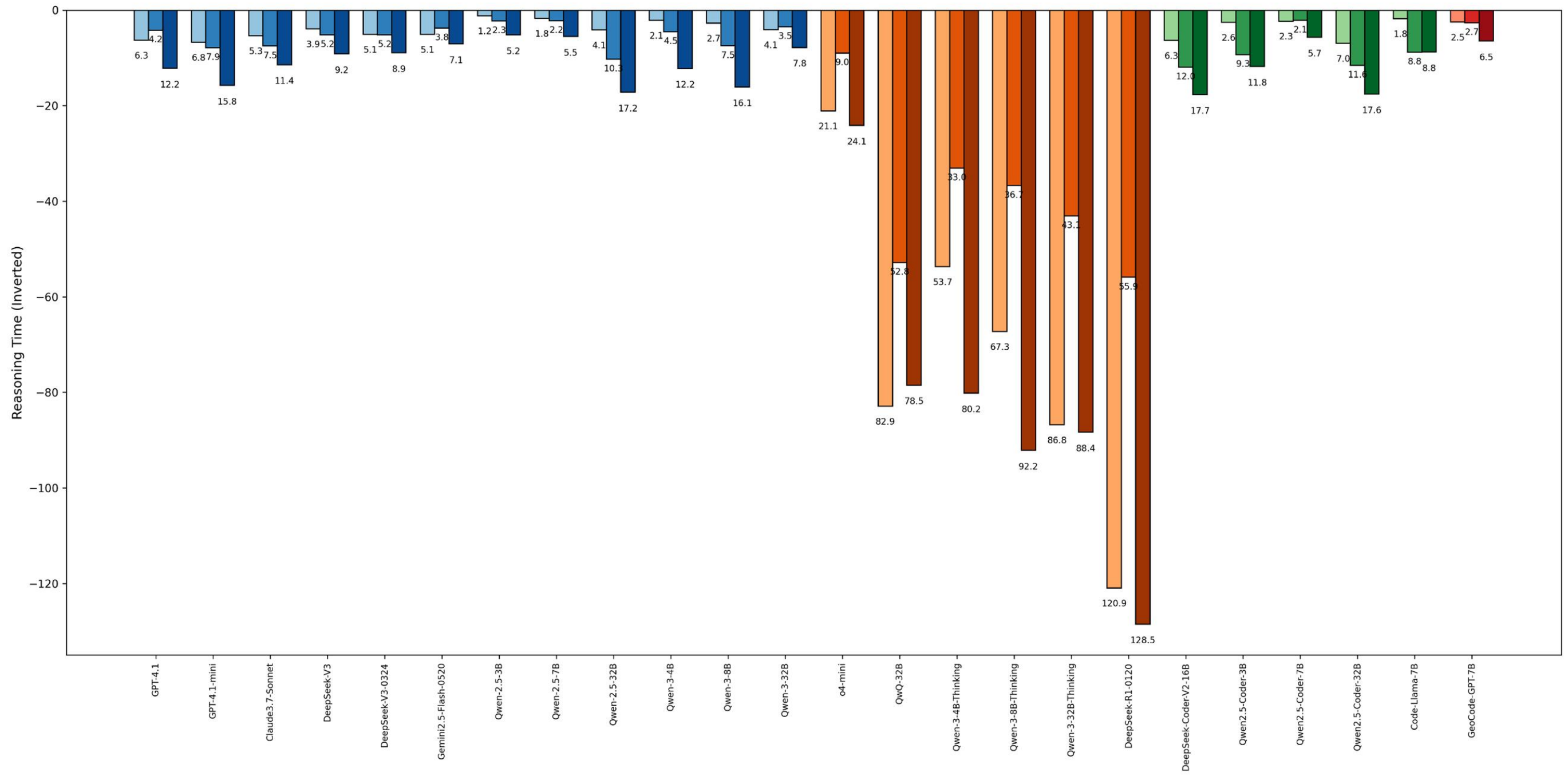

**Figure 21. Reasoning Time Consumption of Models Across Three Testing Tasks.** Colors indicate model categories: blue for general non-reasoning models, orange for general reasoning models, green for general code generation models, and red for geospatial code generation models. Each model has three bars representing unit, integration, and thematic tasks. Bar length reflects reasoning time (inverted display), with longer bars indicating higher time consumption. Absolute values of reasoning time are annotated at the end of each bar.

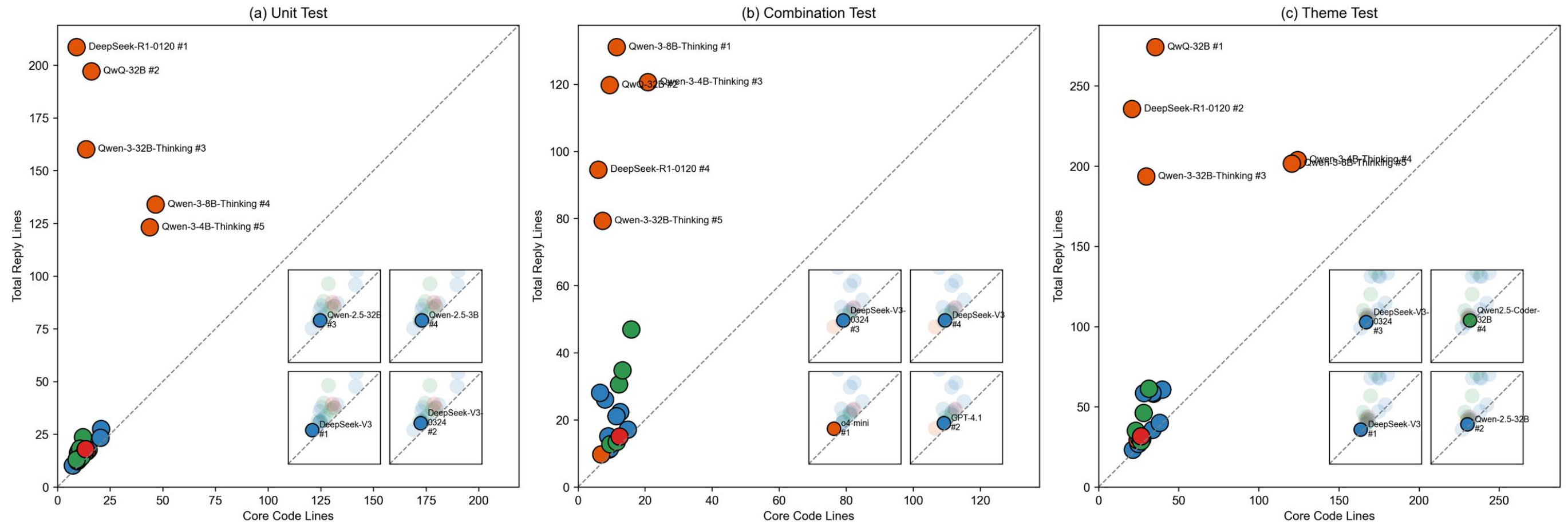

**Figure 22. Scatter plots of code-reply conciseness across three task types.** The scatter plots present the relationship between core code lines (Co.L) and total reply lines (Res.L) for all models in unit test (a), combination test (b), and theme test (c). The grey diagonal line denotes ideal conciseness: points closer to it indicate less redundancy, while proximity to the origin reflects higher efficiency in terms of response length. The five models

with the lowest conciseness are labeled (#1–#5), and the insets in the bottom-right corner highlight the top four models with the highest conciseness in each task.

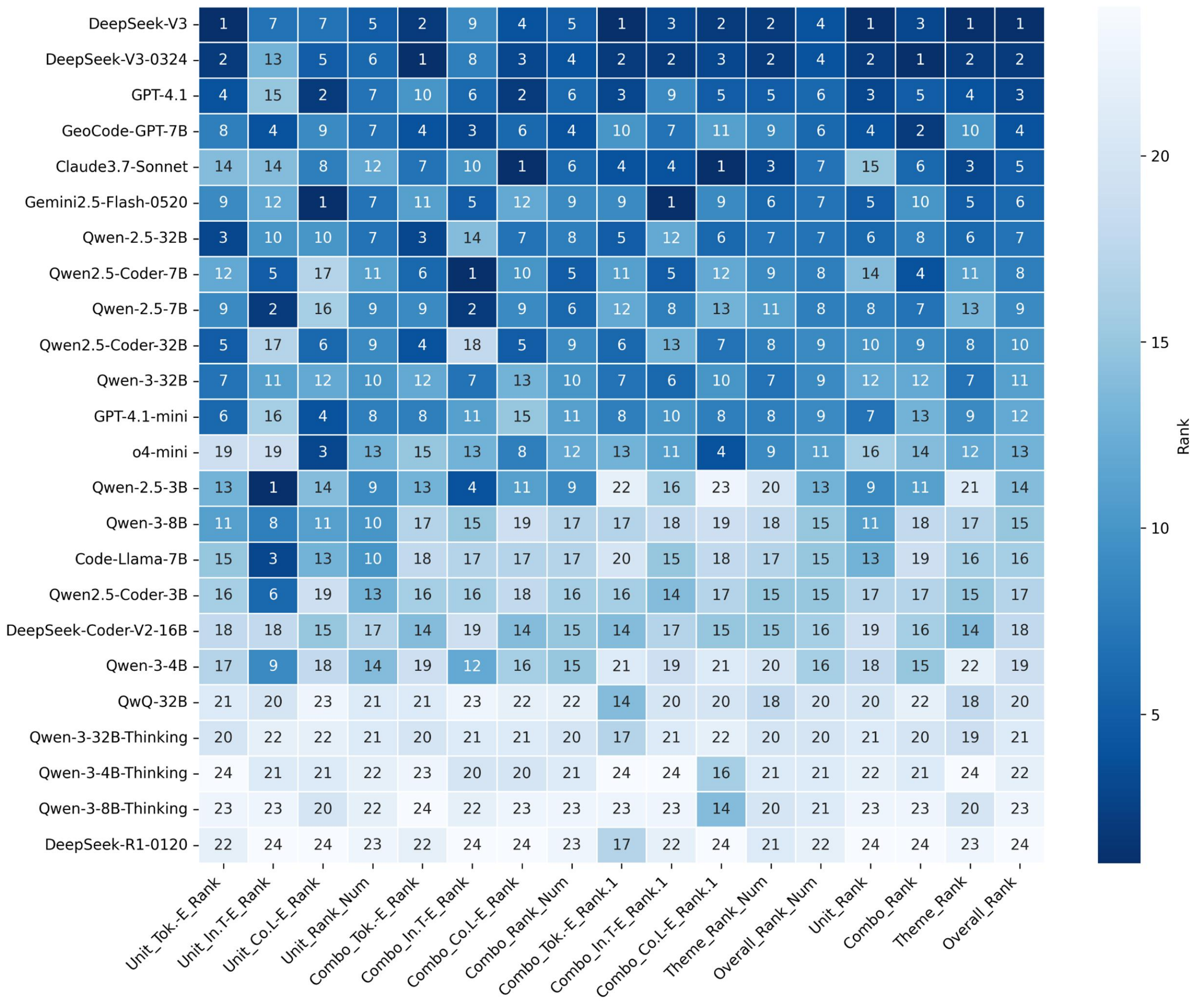

**Figure 23. Heatmap of Model Rankings Across Multiple Tasks.** The heatmap presents model rankings across various task dimensions. Rows represent models, columns denote evaluation criteria, and darker shades indicate better rankings.

## 6.3. Boundary Testing and Error Type Logging

Table 16 lists the average pass rates of each model in the boundary testing tasks. To analyze the relationship between Stability-Adjusted Accuracy (SA) and boundary test Pass Rate, Figure 24 divides the scatter plot into quadrants based on the mean values of both metrics, in order to identify model performance patterns. The figure reveals three distinct distribution characteristics: models in the lower-left quadrant show low performance in both SA and pass rate, mostly consisting of code generation models and small-parameter general models; models in the upper-right quadrant perform well on both metrics and are mainly large-scale commercial general-purpose reasoning models. Notably, the only model located in the lower-right quadrant is the geospatial model GeoCode-GPT-7B, which, despite having a relatively high SA, exhibits a low boundary test pass rate—indicating the need for further optimization under boundary scenarios.

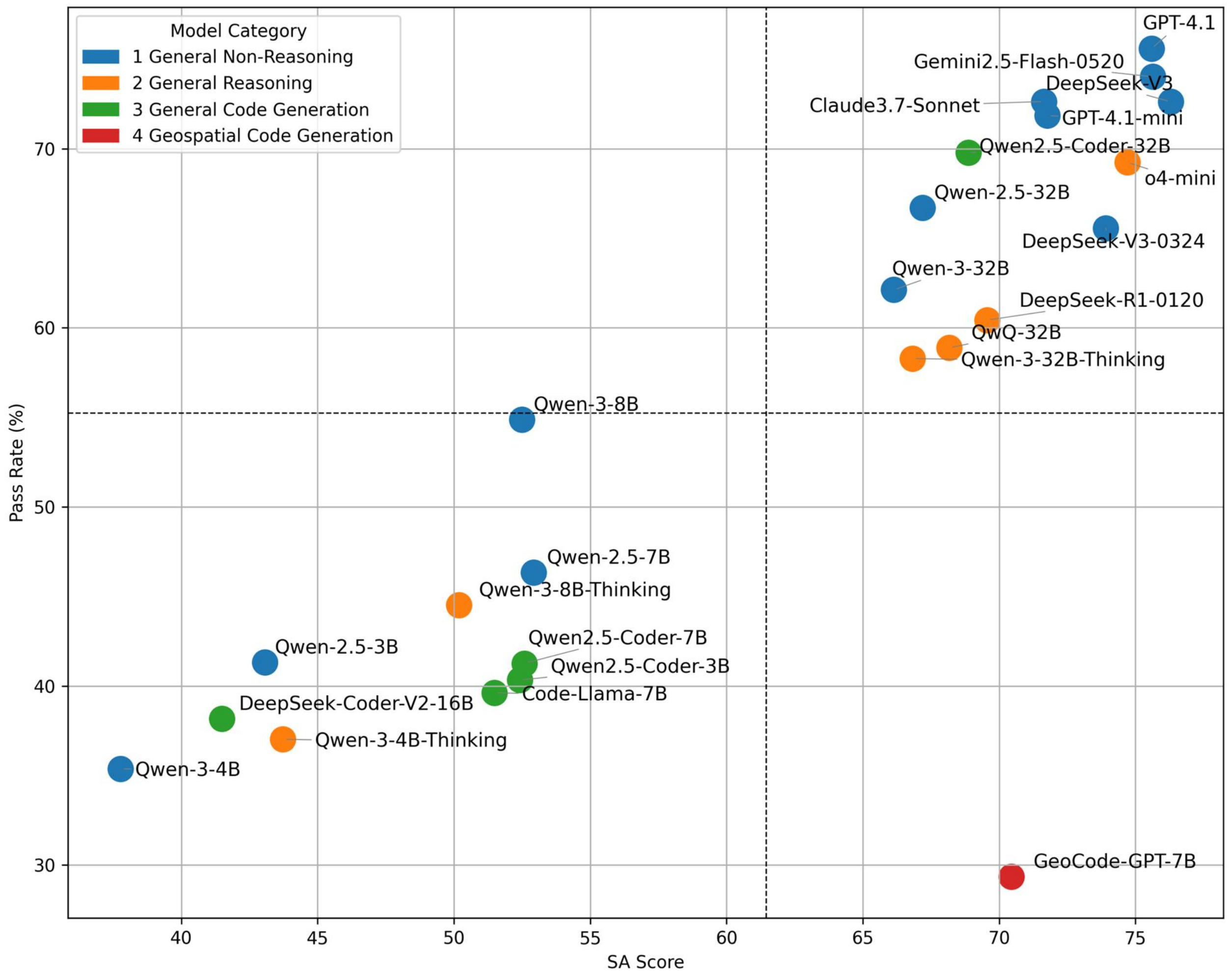

**Figure 24. Quadrant Distribution of Models by Stability and Robustness.** The figure plots each model's stability score (SA) and boundary test pass rate, with mean-based quadrant division to identify performance patterns. Colors indicate model categories; labels denote model names with overlap avoidance.

**Table 16. Accuracy and Boundary Pass Rate of Different Models.**

| Category | Model | Boundary_Pass |
|---|---|---|
| 1 | GPT-4.1 | 75.59 |
| 1 | GPT-4.1-mini | 71.85 |
| 1 | Claude3.7-Sonnet | 72.64 |
| 1 | DeepSeek-V3 | 72.62 |
| 1 | DeepSeek-V3-0324 | 65.55 |
| 1 | Gemini2.5-Flash-0520 | 74.02 |
| 1 | Qwen-2.5-3B | 41.32 |
| 1 | Qwen-2.5-7B | 46.34 |
| 1 | Qwen-2.5-32B | 66.71 |
| 1 | Qwen-3-4B | 35.38 |
| 1 | Qwen-3-8B | 54.87 |
| 1 | Qwen-3-32B | 62.12 |
| 2 | o4-mini | 69.24 |
| 2 | QwQ-32B | 58.88 |
| 2 | Qwen-3-4B-Thinking | 37.02 |

| Category | Model | Boundary_Pass |
| --- | --- | --- |
| 2 | Qwen-3-8B-Thinking | 44.52 |
| 2 | Qwen-3-32B-Thinking | 58.28 |
| 2 | DeepSeek-R1-0120 | 60.43 |
| 3 | DeepSeek-Coder-V2-16B | 38.17 |
| 3 | Qwen2.5-Coder-3B | 40.33 |
| 3 | Qwen2.5-Coder-7B | 41.26 |
| 3 | Qwen2.5-Coder-32B | 69.78 |
| 3 | Code-Llama-7B | 39.62 |
| 4 | GeoCode-GPT-7B | 29.35 |

**Table 17. Error Type Distribution of Each Model on Unit Testing Tasks.**

| Category | Model | Para._Err. | Output_Mis. | Wrong_Ans. | Syn._Err. | Net._Err. |
| --- | --- | --- | --- | --- | --- | --- |
| 1 | GPT-4.1 | 39.72% | 11.25% | 46.26% | 0.08% | 2.68% |
| 1 | GPT-4.1-mini | 42.10% | 8.26% | 45.89% | 1.50% | 2.26% |
| 1 | Claude3.7-Sonnet | 39.96% | 4.47% | 50.10% | 2.68% | 2.80% |
| 1 | DeepSeek-V3 | 65.47% | 6.40% | 24.92% | 0.14% | 3.06% |
| 1 | DeepSeek-V3-0324 | 70.09% | 6.73% | 19.80% | 0.51% | 2.87% |
| 1 | Gemini2.5-Flash-0520 | 47.68% | 7.57% | 40.81% | 1.08% | 2.85% |
| 1 | Qwen-2.5-3B | 77.29% | 7.74% | 10.08% | 4.23% | 0.67% |
| 1 | Qwen-2.5-7B | 78.54% | 3.36% | 13.97% | 2.79% | 1.34% |
| 1 | Qwen-2.5-32B | 63.46% | 8.74% | 23.35% | 2.19% | 2.25% |
| 1 | Qwen-3-4B | 69.85% | 2.08% | 13.72% | 13.58% | 0.77% |
| 1 | Qwen-3-8B | 62.77% | 4.51% | 27.71% | 3.56% | 1.45% |
| 1 | Qwen-3-32B | 52.53% | 8.88% | 35.49% | 0.90% | 2.21% |
| 2 | o4-mini | 54.20% | 5.09% | 37.55% | 0.67% | 2.50% |
| 2 | QwQ-32B | 76.41% | 3.49% | 16.75% | 1.49% | 1.86% |
| 2 | Qwen-3-4B-Thinking | 66.58% | 3.26% | 8.13% | 21.26% | 0.77% |
| 2 | Qwen-3-8B-Thinking | 60.89% | 3.81% | 13.76% | 20.80% | 0.76% |
| 2 | Qwen-3-32B-Thinking | 73.44% | 4.89% | 19.17% | 1.04% | 1.45% |
| 2 | DeepSeek-R1-0120 | 78.99% | 3.32% | 15.06% | 0.35% | 2.27% |
| 3 | DeepSeek-Coder-V2-16B | 77.56% | 4.86% | 12.86% | 3.38% | 1.34% |
| 3 | Qwen2.5-Coder-3B | 73.28% | 9.22% | 11.78% | 4.07% | 1.65% |
| 3 | Qwen2.5-Coder-7B | 82.08% | 4.01% | 11.15% | 0.98% | 1.79% |
| 3 | Qwen2.5-Coder-32B | 45.04% | 20.12% | 31.74% | 0.26% | 2.84% |
| 3 | Code-Llama-7B | 80.02% | 7.76% | 9.97% | 0.97% | 1.28% |
| 4 | GeoCode-GPT-7B | 75.92% | 4.68% | 8.05% | 10.59% | 0.76% |

**Table 18. Error Type Distribution of Each Model on Group Test Tasks.**

| Category | Model | Para._Err. | Output_Mis. | Wrong_Ans. | Syn._Err. | Net._Err. |
| --- | --- | --- | --- | --- | --- | --- |
| 1 | GPT-4.1 | 38.82% | 30.22% | 29.30% | 0.28% | 1.38% |
| 1 | GPT-4.1-mini | 35.19% | 30.72% | 30.82% | 1.89% | 1.39% |
| 1 | Claude3.7-Sonnet | 33.10% | 25.24% | 32.74% | 7.26% | 1.67% |
| 1 | DeepSeek-V3 | 44.94% | 21.97% | 32.18% | 0.00% | 0.91% |

| Category | Model | Para._Err. | Output_Mis. | Wrong_Ans. | Syn._Err. | Net._Err. |
|---|---|---|---|---|---|---|
| 1 | DeepSeek-V3-0324 | 44.91% | 24.84% | 28.69% | 0.27% | 1.28% |
| 1 | Gemini2.5-Flash-0520 | 30.93% | 34.39% | 33.36% | 0.37% | 0.93% |
| 1 | Qwen-2.5-3B | 72.97% | 2.87% | 16.85% | 6.74% | 0.57% |
| 1 | Qwen-2.5-7B | 61.62% | 10.45% | 14.42% | 12.68% | 0.82% |
| 1 | Qwen-2.5-32B | 70.26% | 6.94% | 20.17% | 1.02% | 1.62% |
| 1 | Qwen-3-4B | 71.87% | 6.81% | 7.38% | 13.77% | 0.17% |
| 1 | Qwen-3-8B | 71.76% | 6.00% | 9.74% | 11.96% | 0.55% |
| 1 | Qwen-3-32B | 84.73% | 6.96% | 7.51% | 0.54% | 0.26% |
| 2 | o4-mini | 25.02% | 48.32% | 24.78% | 0.65% | 1.22% |
| 2 | QwQ-32B | 43.09% | 23.60% | 31.91% | 0.49% | 0.90% |
| 2 | Qwen-3-4B-Thinking | 59.89% | 11.88% | 9.66% | 18.26% | 0.31% |
| 2 | Qwen-3-8B-Thinking | 60.63% | 11.39% | 12.74% | 14.63% | 0.61% |
| 2 | Qwen-3-32B-Thinking | 46.33% | 23.53% | 24.71% | 4.85% | 0.59% |
| 2 | DeepSeek-R1-0120 | 35.14% | 35.64% | 27.00% | 0.41% | 1.81% |
| 3 | DeepSeek-Coder-V2-16B | 61.95% | 8.99% | 24.55% | 3.53% | 0.99% |
| 3 | Qwen2.5-Coder-3B | 62.81% | 8.98% | 20.46% | 6.79% | 0.96% |
| 3 | Qwen2.5-Coder-7B | 62.21% | 12.06% | 22.58% | 1.49% | 1.67% |
| 3 | Qwen2.5-Coder-32B | 59.01% | 13.69% | 25.39% | 0.57% | 1.34% |
| 3 | Code-Llama-7B | 72.23% | 5.63% | 19.67% | 1.48% | 1.00% |
| 4 | GeoCode-GPT-7B | 64.35% | 8.39% | 17.00% | 9.38% | 0.88% |

**Table 19. Error Type Distribution of Each Model on Theme Test Tasks.**

| Category | Model | Para._Err. | Output_Mis. | Wrong_Ans. | Syn._Err. | Net._Err. |
|---|---|---|---|---|---|---|
| 1 | GPT-4.1 | 39.93% | 1.12% | 53.73% | 0.00% | 5.22% |
| 1 | GPT-4.1-mini | 40.00% | 1.82% | 48.73% | 4.00% | 5.45% |
| 1 | Claude3.7-Sonnet | 33.18% | 5.00% | 40.00% | 13.64% | 8.18% |
| 1 | DeepSeek-V3 | 51.08% | 3.03% | 39.39% | 0.00% | 6.49% |
| 1 | DeepSeek-V3-0324 | 44.66% | 4.35% | 43.08% | 0.40% | 7.51% |
| 1 | Gemini2.5-Flash-0520 | 53.85% | 0.77% | 38.46% | 0.00% | 6.92% |
| 1 | Qwen-2.5-3B | 81.26% | 0.23% | 5.15% | 12.88% | 0.47% |
| 1 | Qwen-2.5-7B | 81.03% | 2.44% | 12.20% | 3.79% | 0.54% |
| 1 | Qwen-2.5-32B | 70.67% | 1.33% | 23.00% | 0.00% | 5.00% |
| 1 | Qwen-3-4B | 77.48% | 0.73% | 3.15% | 17.68% | 0.97% |
| 1 | Qwen-3-8B | 71.97% | 2.02% | 11.36% | 12.88% | 1.77% |
| 1 | Qwen-3-32B | 65.37% | 3.56% | 27.83% | 0.65% | 2.59% |
| 2 | o4-mini | 40.49% | 0.81% | 50.20% | 1.21% | 7.29% |
| 2 | QwQ-32B | 63.08% | 2.69% | 29.62% | 0.38% | 4.23% |
| 2 | Qwen-3-4B-Thinking | 47.69% | 0.38% | 6.15% | 45.00% | 0.77% |
| 2 | Qwen-3-8B-Thinking | 46.70% | 0.88% | 9.25% | 41.85% | 1.32% |
| 2 | Qwen-3-32B-Thinking | 64.80% | 3.95% | 26.32% | 1.64% | 3.29% |
| 2 | DeepSeek-R1-0120 | 49.22% | 3.12% | 41.02% | 0.00% | 6.64% |
| 3 | DeepSeek-Coder-V2-16B | 75.83% | 0.76% | 16.28% | 5.34% | 1.78% |
| 3 | Qwen2.5-Coder-3B | 79.34% | 2.04% | 9.18% | 7.91% | 1.53% |

| Category | Model | Para._Err. | Output_Mis. | Wrong_Ans. | Syn._Err. | Net._Err. |
|---|---|---|---|---|---|---|
| 3 | Qwen2.5-Coder-7B | 78.47% | 2.72% | 15.26% | 0.82% | 2.72% |
| 3 | Qwen2.5-Coder-32B | 64.85% | 0.68% | 31.06% | 0.00% | 3.41% |
| 3 | Code-Llama-7B | 80.39% | 1.96% | 6.86% | 10.05% | 0.74% |
| 4 | GeoCode-GPT-7B | 81.86% | 1.19% | 3.82% | 12.89% | 0.24% |

We visualized the error performance of each metric across different models, with high and low outliers annotated, as shown in Figure 25. High outliers refer to models with significantly elevated error values, defined as errors greater than or equal to "maximum minus 0.5 times the standard deviation"; low outliers refer to models with significantly low error values, defined as errors less than or equal to "minimum plus 0.5 times the standard deviation." Overall, parameter-related errors are relatively balanced across models; however, it is worth noting that the Qwen-3-Thinking series shows a significantly higher error rate in the syntax error dimension, far exceeding other models, indicating a consistent structural bias.

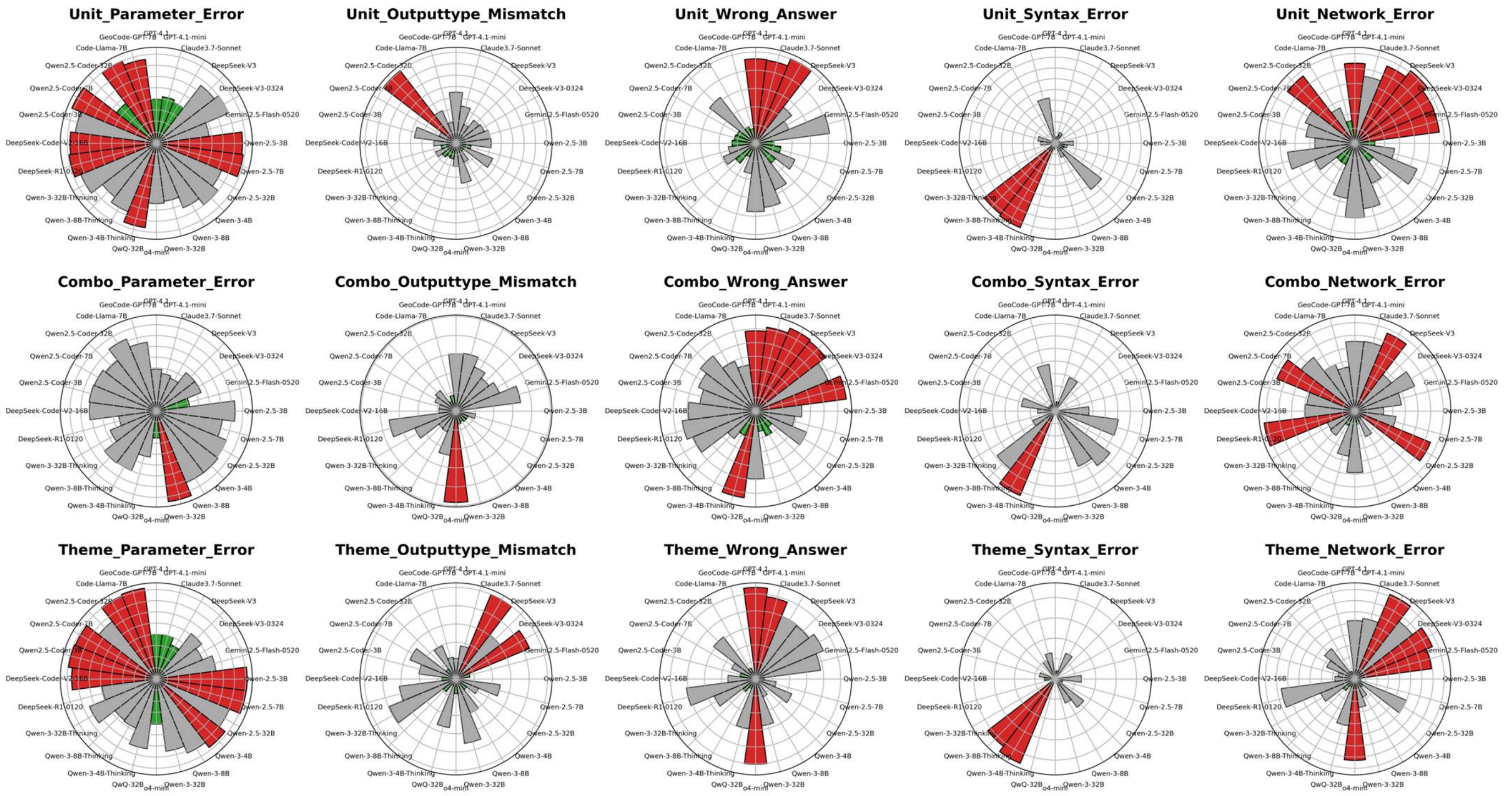

**Figure 25. Error Type Distribution of Each Model.** Each subplot represents one error type; rows correspond to task types (Unit, Combo, Theme), and columns correspond to error categories (Parameter, Output Type Mismatch, Wrong Answer, Syntax, Network). Bar length indicates the error rate per model. Red bars denote high-error outliers, green bars denote low-error outliers, and gray bars indicate normal range.

From the results of the three test types (Figures 26–28), the distribution of error types across major models exhibits clear structural differences and stylistic characteristics. Parameter errors dominate in most models, especially in the Qwen series, where the proportions are significantly higher (e.g., 78.47% in Theme tasks for Qwen2.5-Coder-7B and 84.73% in Group tasks for Qwen-3-32B), suggesting certain limitations in parameter binding and invocation conventions. In contrast, such errors in GPT-4.1 and Claude3.7 remain in the 30–40% range, showing stronger input parsing capabilities. Notably, the Qwen-3-Thinking series shows a significantly higher proportion of syntax errors in Unit and Theme tasks (e.g., 45.00% in Theme for Qwen-3-4B-Thinking), whereas such errors are extremely rare in most models, indicating abnormal fluctuation in language structure control during generation in this series.

Regarding output type errors, o4-mini shows an abnormally high proportion in Group tasks (48.32%, significantly exceeding the ~6% level of most models), possibly reflecting deficiencies in type modeling or output verification mechanisms. There are also individual models, such as Qwen2.5-Coder-32B, that show relatively high type errors in Unit tasks (20.12%), suggesting a possible trade-off in type consistency during complex structure generation.

Logic errors (Answer Errors) account for a higher proportion in some high-performing models. For example, GPT-4.1 reaches 53.73% in Theme tasks, indicating that while it has strong syntax and type control capabilities, there is still room for improvement in result reasoning and semantic alignment. Overall, the Claude and GPT series show a relatively balanced distribution of various error types, while open-source models like Qwen and DeepSeek exhibit concentrated biases in specific types, reflecting differences in model design regarding parameter handling, structure generation, or reasoning control.

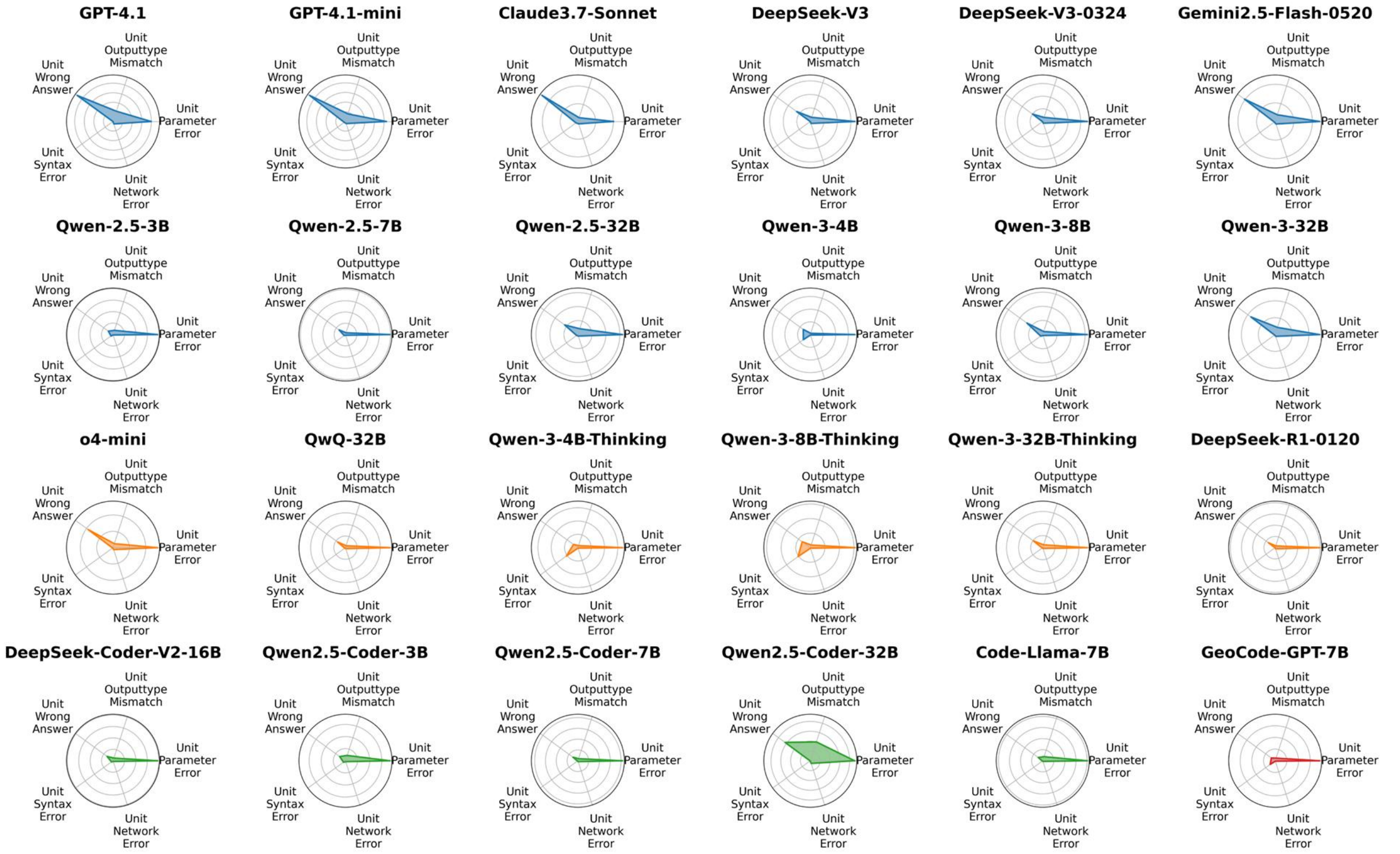

**Figure 26. Unit-Level Error Distribution Across Models.** Each radar chart represents a single model, with axes denoting error types and values indicating their relative proportions in unit-level evaluations.

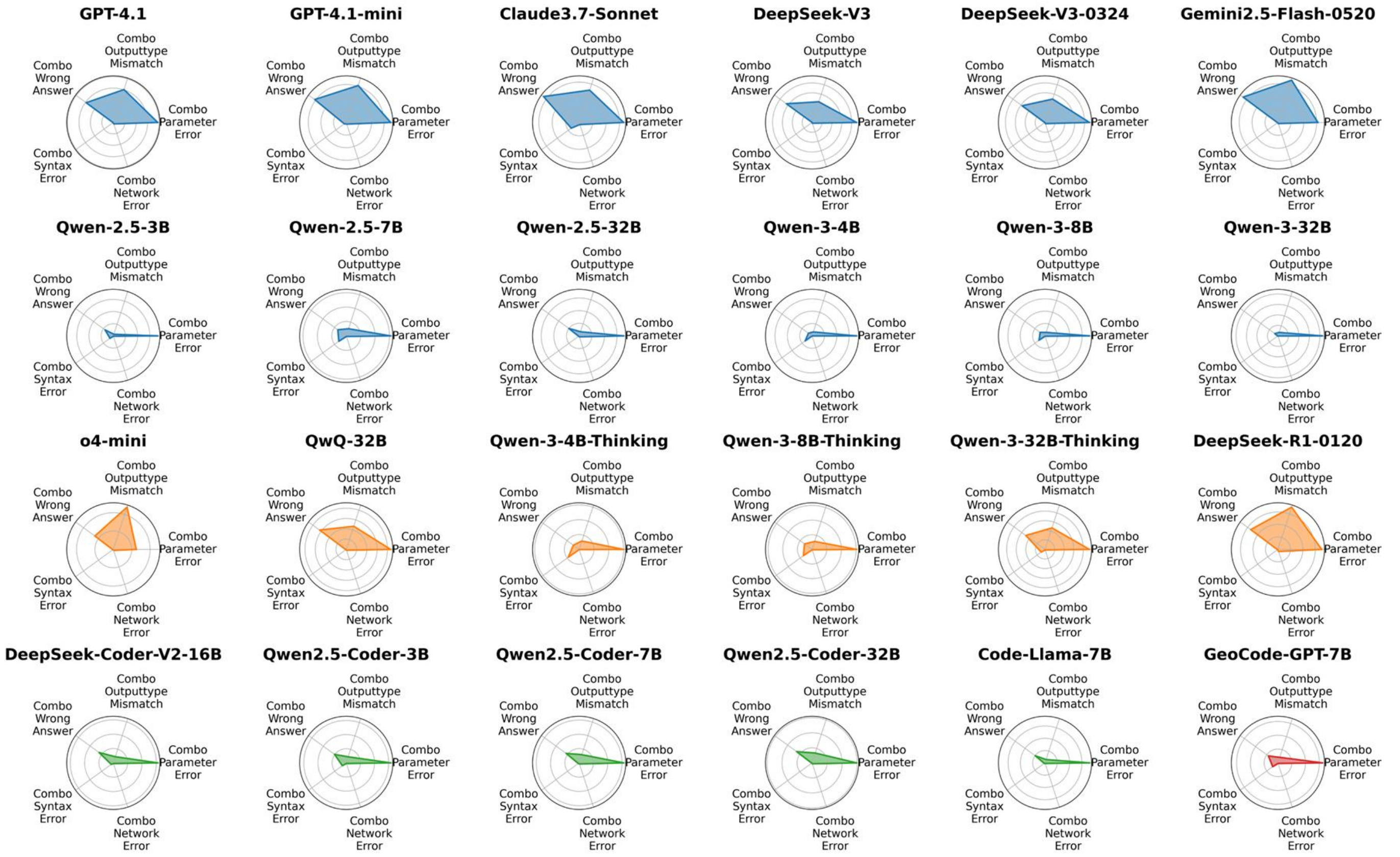

**Figure 27. Combination-Level Error Distribution Across Models.**

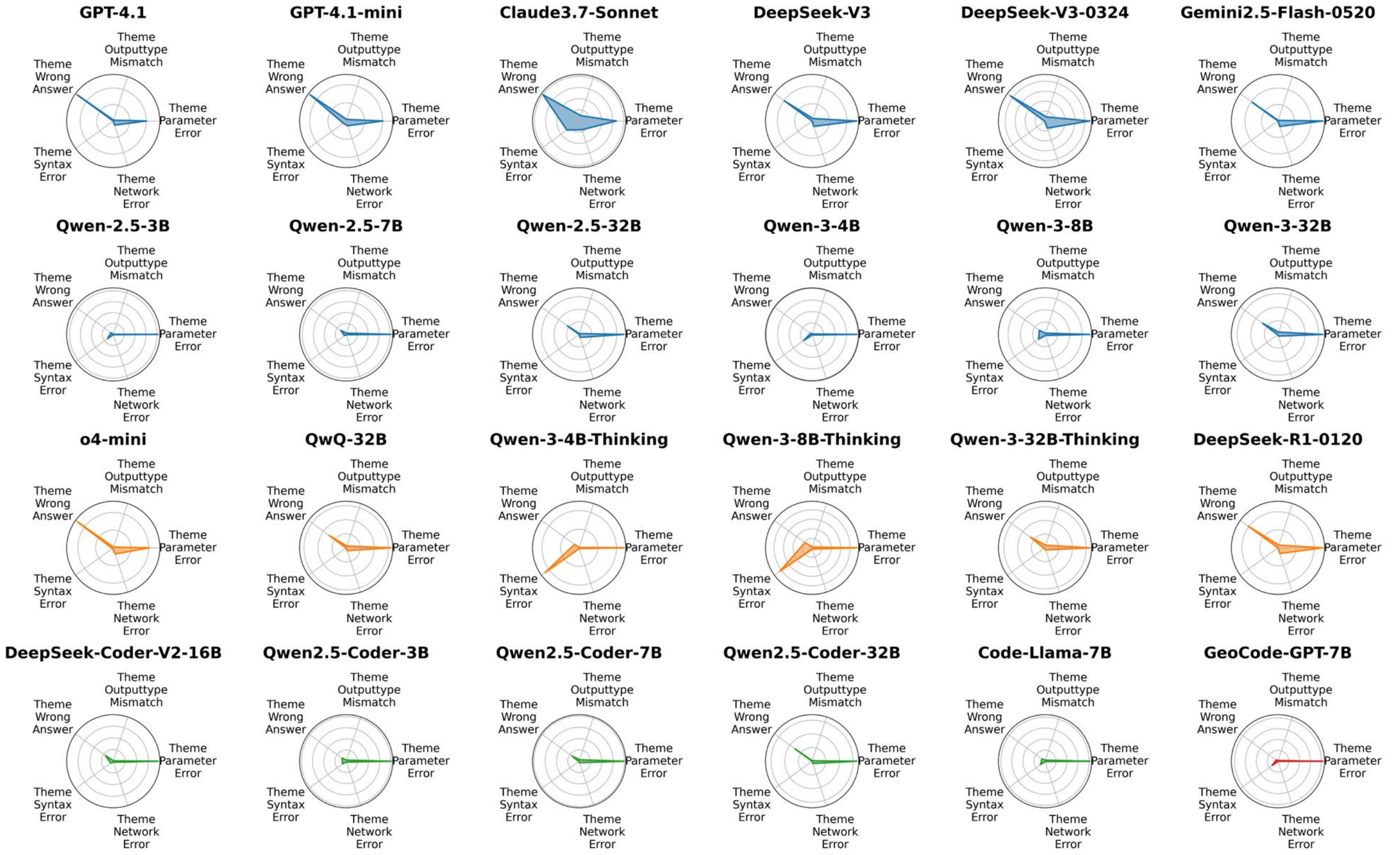

**Figure 28. Theme-Level Error Distribution Across Models.**

## 7. Conclusion

This study proposes AutoGEEval++, the first multimodal and multi-level evaluation framework for LLMs targeting geospatial code generation tasks on the GEE platform. Implemented via the Python interface, the framework supports end-to-end evaluation across 26 GEE data types and multiple levels (unit-level, combination-level, and thematic-level) and modalities. The framework mainly consists of three core components: the constructed test benchmark (AutoGEEval++-Bench), which includes 6,365 test cases comprising 5,078 unit test cases (2,799 general tests and 2,279 boundary tests), 1,199 combination test cases, and 88 thematic test cases; the submission program, which guides LLMs to generate executable code through prompts; and the judge program, which automatically verifies the correctness of model outputs, resource consumption, and error types. Using this framework, we comprehensively evaluated 24 representative LLMs, covering general-purpose models, reasoning-enhanced models, code generation models, and geospatial-specific models.

### 7.1. Core Findings and Insights

Comprehensive experimental results reveal that mainstream LLMs exhibit multidimensional performance differences in geospatial code generation tasks. These differences can be systematically analyzed along five key dimensions: task structure complexity, generation strategy optimization, model architecture type, parameter scale adaptability, and error distribution characteristics.

- **Task Structure Dimension:** Combination tests (averaging 5.14 operators) outperform both unit tests (1 operator) and thematic tests (16.86 operators) in accuracy and generation stability. This suggests that function combinations of moderate complexity better activate structural understanding in models, while highly semantic thematic tasks still pose significant challenges.
- **Generation Strategy Dimension:** Multi-round generation mitigates semantic hallucinations in the initial output and improves overall accuracy—especially from the first to the third generation. However, further increasing the rounds to five shows diminishing returns, highlighting the need to balance redundancy and resource consumption in reasoning strategies.
- **Model Type Dimension:** General-purpose reasoning models show advantages in complex, cross-module thematic tests but incur higher computational costs, limiting their use in resource-sensitive settings. Models like o4-mini achieve a favorable trade-off between accuracy and efficiency, demonstrating strong cost-performance. GPT models display the least fluctuation across all tasks in both accuracy and stability, indicating strong generalization and deployment potential across diverse scenarios.
- **Parameter Scale Dimension:** In local deployment settings, 7B-scale models strike an optimal balance among accuracy, stability, and resource efficiency—outperforming 3B and 32B variants. Notably, GeoCode-GPT-7B ranks fourth overall despite its smaller size, closely following DeepSeek-V3 and GPT-4.1, showcasing robust alignment with geospatial tasks. DeepSeek-V3 leads in accuracy, stability, and efficiency, making it the most capable model overall. However, its specialized Coder variant ranks considerably lower, suggesting a need for improved domain alignment.
- **Error Pattern Dimension:** The most common error types are parameter misusage, type mismatches, and semantic deviations. Syntax errors remain relatively rare, reflecting maturity in structural code generation. Nonetheless, inconsistencies in function invocation, edge condition handling, and domain-specific reasoning remain prominent issues. Code hallucination continues to be a key barrier to model usability and deployment reliability in geospatial applications.

### 7.2. Significance and Advantages

The proposed AutoGEEval++ framework expands the current geospatial code evaluation system across multiple dimensions including task complexity, data diversity, and evaluation granularity. It systematically encompasses unit-level, combination-level, and thematic-level test tasks, significantly enhancing the characterization of LLMs' transferability and generation reliability in geoscientific scenarios. The framework addresses limitations of traditional evaluation methods—such as narrow task types, limited sample sizes, and heavy reliance on manual

judgment—by incorporating an automated error-capturing mechanism and a fine-grained metric system, enabling the detection of code hallucinations and the quantitative analysis of model behavior.

Practically, AutoGEEval++ constructs multimodal test sets targeting boundary conditions, function combinations, and real-world remote sensing tasks, reflecting the high heterogeneity, domain-specificity, and complexity typical of geospatial analysis. This provides a viable pathway for risk screening and deployment optimization prior to model release. Technically, the local runtime implementation based on the GEE Python API ensures evaluation stability, controllability, and reproducibility, overcoming limitations of browser-based environments. For model development, AutoGEEval++ yields feedback for training and fine-tuning geospatially specialized LLMs, supporting the evolution of general-purpose models toward geospatial agents. In summary, the framework not only extends the functionalities of existing tools but also offers a systematic methodology for assessing LLM trustworthiness and domain alignment in the era of GeoAI, with significant academic and engineering value.

### 7.3. Limitations and Future Work

Despite the comprehensive enhancements AutoGEEval++ brings to task types, sample scales, data modalities, and metric systems, several areas remain for future improvement. First, current combination tests are primarily based on structural concatenation and lack complex control logic or real semantic constraints. Future work may introduce task intent-driven mechanisms to enhance logical depth and semantic coherence. Second, thematic tests face trade-offs between coverage breadth and annotation precision; future developments can leverage remote sensing and GIS data platforms to build automated task extraction systems, expanding coverage and structural richness.

On the evaluation side, the current pipeline focuses on single-turn generation and does not yet assess multi-turn dialogue, contextual memory, or incremental debugging—future versions could address this by designing debugging tasks and repair scenarios. Moreover, the current evaluation framework is result-centric and lacks modeling of the generation process itself. Incorporating generation path tracking and sampling behavior analysis could help reveal how the model's "chain of thought" evolves.

Additionally, the root causes of code hallucination and reasoning deviation remain insufficiently understood. Future work could use controlled experiments and ablation studies to analyze hallucination mechanisms from the perspectives of training data, parameter size, and prompt structure. Finally, given the dynamic nature of geospatial tasks, building an open evaluation platform with sustainable updating capabilities—for model submission, automatic scoring, and continuous benchmarking—will be essential for establishing long-term infrastructure for geospatial AI assessment.

## ORCID

Shuyang Hou: https://orcid.org/0009-0000-6984-9959
Huayi Wu: https://orcid.org/0000-0003-3971-0512

## Data availability statement

The experimental data used in this study can be downloaded from https://github.com/szx-0633/AutoGEEval-plus/. Other data that support the findings of this study are available from the corresponding author upon reasonable request.

## Funding

The work was supported by the National Natural Science Foundation of China [Grant number 41971349]